\documentclass[11pt,preprint]{aastex}

\begin{document}

\title{A Deep {\em Hubble Space Telescope} H-Band Imaging Survey of
  Massive Gas-Rich Mergers\footnote{Based on observations with the
    NASA/ESA {\em Hubble Space Telescope}, obtained at the Space
    Telescope Science Institute, which is operated by the Association
    of Universities for Research in Astronomy, Inc. under NASA
    contract No.  NAS5-26555.}}

\author{S. Veilleux and D.-C. Kim}
\affil{Department of Astronomy, University of Maryland, College Park,
MD 20742; \\ veilleux@astro.umd.edu, dckim@astro.umd.edu}

\author{C. Y. Peng}
\affil{Space Telescope Science Institute, 3700 San Martin Drive, Baltimore,
  MD 21218; cyp@stsci.edu}

\author{L. C. Ho} \affil{The Observatories of the Carnegie Institution
  of Washington, 813 Santa Barbara St., Pasadena, CA 91101;
  lho@ociw.edu}

\author{L. J. Tacconi,  K. M. Dasyra, R. Genzel, and D. Lutz}
\affil{Max-Planck-Institut f\"ur extraterrestrische Physik, Postfach
  1312, D-85741 Garching, Germany; linda@mpe.mpg.de, dasyra@mpe.mpg.de, 
  genzel@mpe.mpg.de, lutz@mpe.mpg.de}

\author{D. B. Sanders}
\affil{Institute for Astronomy, University of Hawaii, 2680 Woodlawn
  Drive, Honolulu, HI 96822; sanders@ifa.hawaii.edu}

\begin{abstract}
  We report the results from a deep {\em HST} NICMOS H-band imaging
  survey of a carefully selected sample of 33 luminous, late-stage
  galactic mergers at $z < 0.3$. This program is part of {\em QUEST},
  a comprehensive investigation of the most luminous mergers in the
  nearby universe, the ultraluminous infrared galaxies (ULIRGs) and
  the quasars ({\em QUEST} = {\em Q}uasar / {\em U}LIRG {\em
  E}volutionary {\em S}tudy).  Signs of a recent galactic interaction
  are seen in all of the objects in the {\em HST} sample, including
  all 7 IR-excess Palomar-Green (PG) QSOs in the sample.  Unsuspected
  double nuclei are detected in 5 ULIRGs.  A detailed two-dimensional
  analysis of the surface brightness distributions in these objects
  indicates that the great majority (81\%) of the single-nucleus
  systems show a prominent early-type morphology.  However,
  low-surface-brightness exponential disks are detected on large scale
  in at least 4 of these sources.  The hosts of 'warm' ({\em IRAS}
  25-to-60 $\mu$m flux ratio, $f_{25}/f_{60} > 0.2$), AGN-like systems
  are of early type and have less pronounced merger-induced
  morphological anomalies than the hosts of cool systems with LINER or
  HII region-like nuclear optical spectral types.  The host sizes and
  luminosities of the 7 PG~QSOs in our sample are statistically
  indistinguishable from those of the ULIRG hosts.  In comparison,
  highly luminous quasars, such as those studied by Dunlop et
  al. (2003), have hosts which are larger and more luminous. The hosts
  of ULIRGs and PG~QSOs lie close to the locations of
  intermediate-size ($\sim$ 1 -- 2 $L^*$) spheroids in the photometric
  projection of the fundamental plane of ellipticals, although there
  is a tendency in our sample for the ULIRGs with small hosts to be
  brighter than normal spheroids.  Excess emission from a young
  stellar population in the ULIRG/QSO hosts may be at the origin of
  this difference. Our results provide support for a possible
  merger-driven evolutionary connection between cool ULIRGs, warm
  ULIRGs, and PG~QSOs. However, this sequence may break down at low
  luminosity since the lowest luminosity PG~QSOs in our sample show
  distinct disk components which preclude major (1:1 -- 2:1) mergers.
  The black hole masses derived from the galaxy host luminosities
  imply sub-Eddington accretion rates for all objects in the sample.
\end{abstract}


\section{Introduction}

Galaxy merging is a key driving force of galaxy evolution. In
hierarchical cold dark matter models of galaxy formation and
evolution, merging leads to the formation of elliptical galaxies,
triggers major starbursts, and may account for the growth of
supermassive black holes and the formation of quasars (e.g., Kauffmann
\& Haehnelt 2000). The importance of mergers increases with redshift
(e.g., Zepf \& Koo 1989; Carlberg, Pritchet, \& Infante 1994;
Neuschaefer et al. 1997; Khochfar \& Burkert 2001). It is clear that
dust-enshrouded starbursts and active galactic nuclei (AGN) play an
extremely important role in the high-redshift Universe and are
probably the dominant contributors to the far-infrared/submm and X-ray
backgrounds, respectively (e.g., Pei, Fall, \& Hauser 1999; Miyaji,
Hasinger, \& Schmidt 2000).  These luminous, merger-induced starbursts
and AGN at high redshift thus provide readily observable signposts for
tracing out the main epoch of elliptical galaxy and quasar formation
if the above scenario is correct.

In order to assess quantitatively the physics of the merger process
and its link to the epoch of elliptical and QSO formation at high
redshift we must first understand the details of galaxy merging and
its relationship to starbursts and AGN in the local universe. The most
violent local mergers and the probable analogs to luminous
high-redshift mergers are the ultraluminous infrared galaxies
(ULIRGs). ULIRGs are advanced mergers of gas-rich, disk galaxies
sampling the entire Toomre merger sequence beyond the first
peri-passage (Veilleux, Kim, \& Sanders 2002; hereafter referred as
VKS02). ULIRGs are among the most luminous objects in the local
universe, with both their luminosities ($>$ 10$^{12}$ L$_\odot$
emerging mainly in the far-IR) and space densities similar to those of
quasars (e.g., Sanders \& Mirabel 1996). The near-infrared light
distributions in many ULIRGs appear to fit a $R^{1/4}$ law (Scoville
et al. 2000; VKS02).  ULIRGs have a large molecular gas concentration in
their central kpc regions (e.g., Downes \& Solomon 1998) with
densities comparable to stellar densities in ellipticals. These large
central gas concentrations (and stars efficiently forming from them)
may be the key ingredient for overcoming the fundamental phase space
density constraints that would otherwise prevent formation of dense
ellipticals from much lower density disk systems (Gunn 1987;
Hernquist, Spergel, \& Heyl 1993). Kormendy \& Sanders (1992) have
proposed that ULIRGs evolve into ellipticals through merger induced
dissipative collapse.  In this scenario, these mergers first go
through a luminous starburst phase, followed by a dust-enshrouded AGN
phase, and finally evolve into optically bright, `naked' QSOs once
they either consume or shed their shells of gas and dust (Sanders et
al. 1988a).

Gradual changes in the far-infrared spectral energy distributions
between `cool' ULIRGs ({\em IRAS} 25-to-60 $\mu$m flux ratio,
$f_{25}/f_{60} < 0.2$), `warm' ULIRGs, and QSOs (Sanders et al. 1988b;
Haas et al. 2003) bring qualitative support to an evolutionary
connection between these various classes of objects, but key elements
remain to be tested. In a pilot study of a dozen ULIRGs observed with
Keck and VLT, Genzel et al. (2001) and Tacconi et al. (2002) have
found that ULIRGs resemble intermediate mass ellipticals/lenticulars
with moderate rotation, in their velocity dispersion distribution,
their location in the fundamental plane (e.g., Djorgovski \& Davis
1987; Dressler et al. 1987; Kormendy \& Djorgovski 1989) and their
distribution of the ratio of rotation/velocity dispersion [v$_{\rm
rot}$ sin(i)/$\sigma$]. These preliminary results therefore suggest
that ULIRGs form moderate mass ($m^\ast \sim 10^{11}$ M$_\odot$), but
not giant (5 -- 10 $\times$ 10$^{11}$ M$_\odot$) ellipticals. A
comparison between these ULIRGs and the sample of luminous radio-quiet
QSOs from Dunlop et al. (2003) indicates that the ULIRGs are offset
from the location of the hosts of these QSOs in the photometric
projection of the fundamental plane.
The latter fall near the locale of giant ellipticals
on the fundamental plane. The black hole (BH) masses inferred for the
ULIRGs from the host dynamical masses and the local BH mass to bulge
velocity dispersion relationship (e.g., Gebhardt et al. 2000) are akin
to those of local Seyfert galaxies. From this perspective, ULIRGs and
QSOs have comparable luminosities because the ULIRGs are much more
efficiently forming stars and feeding their BHs at the epoch they are
observed.

This pilot study (Genzel et al. 2001; Tacconi et al. 2002) has raised
important questions towards understanding the evolution of ULIRG
mergers and their relation to QSOs, but is limited by the small number
of galaxies. For example, giant ellipticals constitute only $\sim$ 10
-- 15\% of all ellipticals with mass $>$ 10$^{11}$ M$_\odot$ -- a
similar fraction of the most massive ULIRG hosts might have been
missed in our small pilot sample. To better understand the
evolutionary process of ULIRGs and their relation to elliptical
galaxies and quasars, we are currently conducting a comprehensive
study of a large sample of ULIRGs and QSOs which samples the full
range of properties (L$_{\rm IR}$, merger state, and AGN/starburst
fraction). This study is called {\em QUEST} (for {\em Q}uasar/{\em
U}LIRG {\em E}volutionary {\em ST}udy).  Our program relies on {\em
HST} NICMOS imaging and VLT/Keck near-infrared spectroscopy to
determine the structure and dynamics of ULIRG and QSO mergers, and on
high-S/N {\em Spitzer Space Telescope} ({\em SST}) mid-infrared
spectroscopy to quantify the energy production mechanisms in these
objects, and to study their obscuration and physical conditions along
the merger sequence.

The purpose of the present paper is to report the results from the
{\em HST} NICMOS imaging component of {\em QUEST}. Part of the
spectroscopy has been presented in Dasyra et al. (2005). In \S 2, we
describe the {\em HST} sample. Next we discuss the observational
strategy of our program and the methods used to obtain, reduce, and
analyze the data (\S 3, \S 4, and \S 5, respectively). The results are
presented in \S 6 and discussed in \S 7. The main conclusions are
summarized in \S 8.  Throughout this paper, we adopt $H_0$ = 75 km
s$^{-1}$ Mpc$^{-1}$ and $q_0$ = 0 to be consistent with previous
papers in this series.  Given the proximity ($z < 0.27$) of all of our
objects, adoption of the WMAP cosmological parameters ($\Omega_M =
0.27$, $\Omega_\Lambda$ = 0.73, H$_0$ = 71 km s$^{-1}$ Mpc$^{-1}$)
would not affect any of our conclusions.

\section{HST Sample}

The ULIRG component of {\em QUEST} focuses on the 1-Jy sample, a
complete flux-limited sample of 118 ULIRGs selected at 60 $\mu$m from
a redshift survey of the IRAS faint source catalog (Kim \& Sanders
1998). 
The QSO component focusses on the Palomar-Green (PG) quasars of
Schmidt \& Green (1983).  The 1-Jy ULIRGs and PG~QSOs are well matched
in bolometric luminosity and redshift ($z < 0.3$). Ground-based
optical and near-infrared images are available for all objects in the
{\em QUEST} sample (e.g, Surace \& Sanders 1999; Surace et al. 2001;
Kim, Veilleux, \& Sanders 2002; VKS02).
However, many of these images, especially those from Kim et al. (2002)
and VKS02, were obtained under seeing $\sim$ 0$\farcs$5 -- 1$\farcs$0,
or $\sim$ 1 -- 3 kpc at $z \sim 0.15$, so the structural parameters of
highly nucleated late-stage ULIRGS (i.e.  objects with bright AGN or
nuclear starbursts) derived from these data are uncertain.  The main
rationale behind our {\em HST} NICMOS program is to revisit these
particular objects. The {\em HST} sample consists of 26 highly nucleated
1-Jy ULIRGs and 7 PG~QSOs.\footnote{ Mrk~1014 is both a 1-Jy ULIRG and
a PG~QSO. In the present paper, we adopt the QSO classification for
this object. }  Table 1 lists some of the properties of these systems.
We have VLT/Keck kinematic data for most of these objects. 
The ULIRGs in the {\em HST} sample were selected to have high nuclear
concentration indices (ratio of the H-band luminosity from the inner 4
kpc to the total luminosity larger than $\sim$ 1/3; VKS02) and the
majority of them host strong, if not dominant, optical and infrared
AGNs (Veilleux, Kim, \& Sanders 1999a; Veilleux, Sanders, \& Kim
1999b; Lutz, Veilleux, \& Genzel 1999 and references therein).  This
subset of ULIRGs is therefore AGN biased and not representative of the
1-Jy sample as a whole. The 7 PG~QSOs in the sample have IR-to-optical
luminosity ratios which are similar to those of the 1-Jy ULIRGs, {\em
i.e.} they are IR-excess QSOs with L$_{\rm IR}$/L$_{\rm BOL}$ $>$
0.4. They were selected from the sample of Surace et al. (2001) and
are probably not representative of the PG~QSO sample in general.  They
represent a small subset of the {\em QUEST} QSO sample. Two of the
QSOs in the {\em HST} sample have absolute B-band magnitudes which are
significantly fainter than the traditional luminosity threshold of
QSOs ($M_B = -23$ for $H_0 = 50$ km s$^{-1}$ Mpc$^{-1}$; see Table 1).

\section{Observational Strategy and Data Acquisition}

The main observational goals of our program are to extract the central
point sources from our targets and derive accurate structural
parameters of the underlying hosts.  The excellent spatial resolution and
sensitivity of the {\em HST} NICMOS camera in the non-thermal infrared
are best suited for these observations.  NICMOS was selected over ACS
to reduce the impact of dust extinction and star formation on the
measurements (especially in the cores of ULIRGs; VKS02). The strong
thermal background of NICMOS makes deep observations at K unrealistic;
our program therefore focuses on the H band (F160W filter), roughly
matching the waveband of our Keck and VLT spectra.  Given the
redshifts of our targets (z $\sim$ 0.05 -- 0.25) and the strengths of
the emission features in these objects (see, e.g., Veilleux et al.
1999b; Genzel et al. 2001; Tacconi et al.  2002), contamination by
emission lines (mainly [Fe~II] and Pa$\beta$) is at most $\sim$ 10\%
for the F160W filter, and is therefore not an issue here.

The need for deep images can hardly be overstated.  Comparisons of our
ground-based R (6400 \AA) and K$^\prime$ (2.1 $\mu$m) imaging of
ULIRGs with the results derived from shallow (e.g., SNAP) {\em HST}
images show that too shallow {\em HST} data underestimate the
luminosities and half-light radii of the hosts, make profile fitting
ambiguous (e.g., de Vaucouleurs versus exponential disk), and can even
completely overlook low-surface-brightness exponential disks extending
significantly beyond galactic bulges (e.g., see discussion in \S 4 in
VKS02; also Peng et al. 2002).  Our ground-based analysis of the 1-Jy
sources shows that we need to reach detection levels of about $\sim$
21.5 mag.~arcsec$^{-2}$ at K$^\prime$ (i.e.  about 1.5
mag.~arcsec$^{-2}$ deeper than the data of VKS02) to avoid these
problems.  This is equivalent to $\sim$ 21.8 mag.~arcsec$^{-2}$ at H
assuming colors that are typical of elliptical galaxies at z = 0.1 --
0.2 (e.g., Lilly \& Longair 1984).

The NIC2 camera was chosen for this program, based on the requirements
of good sensitivity to low surface brightness features, excellent
spatial resolution (0$\farcs$076 pixel$^{-1}$) for accurate PSF (FWHM
= 0$\farcs$14) removal, and a field of view (19$\farcs$5 $\times$
19$\farcs$5) large enough to encompass most of the structures in our
targets.  With this camera, the aforementioned H-band flux level is
reached in $\sim$ 45 minutes with the F160W filter. One full orbit was
therefore used to acquire deep images of each target, for a total of
33 orbits.  A centered 4-point spiral dither pattern
(NIC-SPIRAL-DITHER) with steps of 22.5 pixels was used to better
sample the instrumental PSF, and aid with the recognition and
elimination of data artifacts (e.g., dead and hot pixels, unstable
columns, cosmic-ray afterglow). A shorter sequence of exposures was
taken at the beginning of each orbit to make full use of the orbit and
reduce the impact of persistence; this sequence was not used in the
final analysis. At each dither position, a logarithmic STEP64
MULTIACCUM sequence with NSAMP = 18 (except for the first sequence
where NSAMP = 8-12) was used to provide the largest dynamic range and
allow the calibration to recover the bright central point source. The
resulting exposure time for each object is 2560 seconds.  An
additional orbit was used to obtain several MULTIACCUM exposure
sequences of two stars (SA 107-626 and SA 107-627) and fully
characterize the PSF.  We find below (\S 5) that the stellar PSF
derived from these data is adequate to remove the central point source
from the objects in our sample.

\section{Data  Reduction} 

The raw {\em HST} NICMOS data were first processed with the IDL
procedure {\em undopuft.pro} written by Eddie Bergeron at the STScI
for removal of electronic echoes of bright source and associated
stripes (known as ``Mr. Stay-Puft'').  These data were subsequently
processed with the standard pipeline processing task {\em calnica}
within IRAF/STSDAS.  This task corrects for the non-linearity of the
detector and removes bias value, dark current, amplifier glow, and
shading.  Variable quadrant bias or ``pedestal'' that cannot be
removed by {\em calnica} was removed using the {\em pedsky} task.
This task also removes the sky signal from each image.  A visual
inspection of the data shows that this method gives excellent results
even for objects which cover more than half of the field of view
(F05189--2524 may be the only problem case; see \S 6). Data taken
after the South Atlantic Anomaly (SAA) passage show cosmic ray
persistence that decays with time and leaves a mottled, blotchy or
streaky pattern of noise signal across the images.  The IDL procedure
{\em saa\_clean.pro} was used to remove this effect.  Next, the four
dithered exposures of each object were combined using the ``drizzle''
technique (Gonzaga et al.  1998).  This technique combines dithered
images while preserving photometric accuracy, enhancing resolution,
and correcting geometric distortion. In the same process, this
technique removes cosmic rays, the central bad column, and the
coronographic hole. Warm pixels are not noticeable in our data. For
the photometric calibration of the reduced data, a Vega-normalized
magnitude for F160W (NIC2) was derived following the recipe in the
{\em HST} Data Handbook for NICMOS (Dickinson et al.  2002) using the
calibration appropriate for Cycle
12 data obtained with warm, T = 77.1 K, detectors: $m$(F160W) = $-$2.5
log [PHOTFNU $\times$ CR / $f_\nu$ (Vega)] = 22.107 $-$ 2.5 log (CR)
where PHOTFNU = 1.49816 $\times$ 10$^{-6}$ Jy sec DN$^{-1}$, a keyword
in the calibrated data which indicates the bandpass-averaged flux
density for a source that would produce a count rate of 1 DN
sec$^{-1}$, CR is the count rate in DN sec$^{-1}$, and $f_\nu$(Vega) =
1043.5 Jy, the flux density of Vega in the F160W band for NIC2.

\section{Data Analysis}

The two-dimensional fitting algorithm GALFIT (Peng et al. 2002) was
used to accurately remove the central point source in each object and
determine the structural parameters of the underlying host. The
algorithm improves on previous techniques in two areas, by being able
to simultaneously fit one or more galaxies with an arbitrary number of
components (e.g., S\'ersic profile, exponential disk, Gaussian or
Moffat functions), and with optimization in computation speed. The
azimuthal shapes are generalized ellipses that can fit disky or boxy
components.

In galaxy fitting it is very much a judgment call when it comes to
deciding which types of components and how many components to fit.
Because the studies in literature ({\em i.e.} our comparison sample)
were mostly based on a single component S\'ersic profile, or in some
cases using two ({\em i.e.} bulge-disk decomposition), to make a
sensible comparison, we have to follow the same strategy.  Thus, the
technique we settled on for the ULIRG/PG~QSO project was motivated
by interests to: (1) compare the gross, {\em i.e.} average, galaxy
morphology of ULIRGs and PG~QSOs to the early-type galaxy population,
and (2) in ULIRGs or PG~QSOs where there appeared to be two
components, to do a bulge-disk decomposition to isolate just the bulge
for a more direct comparison with early-type galaxy properties.  This
meant that we did not fit all galaxies to very high detail, e.g. tidal
features are fitted by unrealistic ellipsoid components.

In contrast, in Peng et al. (2002) one goal was to model galaxy
profiles in detail to see whether there were sub-components embedded
inside early-type galaxies.  The other goal was to illustrate the
capability of GALFIT for doing multi-component fitting.  Therefore the
goals called for using as many components as necessary to do a highly
faithful decomposition, in terms of reducing the residuals down to
near the noise level, even if some of the components may not be real.
In our ULIRG/QSO study, we are interested only in physically meaningful
components, as we understand them.  But because of the complexity, we
also must use quite a bit of judgment and intuition to tell us what
``physically meaningful'' means, instead of trying to fit everything we
see in an image.  However subjective, admittedly, this is the only
recourse. 

\subsection{One Galaxy Component Fits}

The analysis of each object followed a number of well-defined steps.
First, we constructed a mask to exclude bright stars or small
foreground/background galaxies within the field of view. Next, we
proceeded to fit the surface brightness distribution of each object
using a single S\'ersic component (observed intensity profile $I
\propto {\rm exp}[-R^{1/n}]$) to simulate the galaxy host and a PSF
model to account for the possibility of an unresolved nuclear
starburst or AGN. The high-S/N PSF model was derived from our deep
images of SA 107-626 as described in \S 3 above.  SA 107-627 was found
to show a suspicious shoulder near the profile core and was not used
to create the PSF model. We found that this method gave virtually
identical results to the use of theoretical Tiny Tim PSFs. 

The S\'ersic component was convolved with the PSF before comparison
with the data. Three S\'ersic components were examined: $n$ = free
({\em i.e.}  left unconstrained), $n$ = 1 (exponential disk profile),
and $n$ = 4 (de Vaucouleurs profile). In all cases, the centroids of
the PSF and S\'ersic components were left unconstrained. This
relatively simple one galaxy component analysis allowed us to get a
general sense of the complexity of each system and whether the system
is disk- or spheroid-dominated.

\subsection{Two Galaxy Component Fits}

The residuals from the one component galaxy fit are often quite
significant.  This is generally the results of merger-induced
morphological anomalies. However, in other cases, these residuals may
indicate the presence of a second low-surface-brightness galaxy
component (e.g., disk). So we decided to look into this possibility by
adding a second (PSF-convolved) galaxy component to the fits for each
object and examining the effects on the goodness of the fits. We
studied the following three cases: ($n$ = 1) + ($n$ = free); ($n$ = 4)
+ ($n$ = free); ($n$ = 1) + ($n$ = 4). Here again, the centroids of
the various components were left unconstrained. Not surprisingly given
the larger number of free parameters, these two-component models
generally provide better fits to the data.  However, a careful
examination of the fitted components often indicate that the second
galaxy component is not physically meaningful.

There are usually many ways to recognize when a second galaxy
component is not real or not physically meaningful.  For instance, the
most common symptoms of unrealistic components are when one or more
parameters converge to extreme values such as: 1. Luminosity becomes
too faint (implying the component is not needed), or too bright
(implying the S\'ersic index is too high). 2. Size becomes too small
(implying PSF artifacts) or extremely large (implying sky problem or
neighboring contamination). 3. Axis ratio is too small, implying PSF
artifact. 4. Position mis-centered, implying peculiar morphology
(e.g., merger-induced morphological anomalies) or the component has
gone and fitted one of the neighbors. 5. S\'ersic index $n$ is large
($n > 4$), implying 2 or more components, PSF artifacts, neighboring
contamination, sky problems, or a combination of some of these
effects.  Condition 5 is usually the telltale sign for multiple
components.  Based on experience, such systems can sometimes have
bulge+disk components, if the PSF residuals are not the dominant cause
for a high $n$.  Sometimes, having a large S\'ersic index also
indicates there is contamination due to neighboring objects -- in
which case both galaxies should be fitted simultaneously to better
determine the morphology of the primary object.

The surest way to recognize when a second galaxy component is real is
to ``put back'' into the residual image the model components
individually to see which structure was being fitted.  The components
have to be fairly distinct both spatially (e.g. axis ratio, size,
centering) and morphologically (concentration) for us to accept the
two components as being real in our assessment.

\subsection{Binary Systems}

Finally, in five systems, the residuals from both the one and
two galaxy component analyses show the presence of a previously
unsuspected companion galaxy separated from the main component by less
than 1\arcsec. In these cases, we limit our analysis to a
one galaxy component analysis (PSF + S\'ersic component) for {\em each}
galaxy in the system, leaving the centroids and S\'ersic indices
unconstrained.

\subsection{Independent Check on the Analysis}

In some cases, the analysis was carried out a second time by other
members of our group to independently verify the significance of the
results.  There are three aspects of the analysis that require some
subjective judgment.  As mentioned above, we must decide on the number
of components to use and their physical realities: we do this by
fitting multiple components, and examining the parameters to determine
whether they are significant galaxy components that perhaps represent,
for example, bulge, disk, bar, compact nuclear component.  In
addition, some PSF residuals due to PSF mismatch may in fact be large,
thus one must decide whether the galaxy component is significantly
affected (and if so, what to do about it: masking or fitting).
Thirdly, in systems that appear to be highly interacting or have large
tidal tails, one has to decide how to perform a decomposition: do we
mask out the neighbor or fit them together, representing the tidal
feature by another component.  Thus this independent
reanalysis by different members of our team addresses particular
subtle cases that fall under one or some of these categories.

\section{Results}

The results from the GALFIT analysis are shown in Figures 1 -- 3 and
listed in Tables 2 -- 4.  Figure 1 presents the residuals found after
subtracting one galaxy component models (PSF + S\'ersic with $n$ =
free, 1, or 4) from the surface brightness distributions of
single-nucleus systems in our sample. Also shown in this figure are
the azimuthally-averaged radial intensity profiles for each scenario
($n$ = free, 1, or 4) and the data.  In a few cases (discussed in more
detail in \S 6.2), we find that adding another S\'ersic component
significantly improves the goodness of the fits; the results of this
more sophisticated two galaxy component analysis are shown in Figure
2. The structural parameters derived from the one and two galaxy
component fits are listed in Tables 2 and 3, respectively. Note that
the exact value of $n > 4$ is not too significant.  It generally
indicates the galaxy has either a strong core (e.g., bulge dominated)
or an extended wing (e.g. early-type galaxies or
interacting/neighboring galaxies), or both.  Large $n$ can be caused
by bad AGN subtraction, but we tried to minimize that likelihood by
using multiple components for the core.  We also tried to minimize
neighboring contamination by fitting the neighbors and/or masking.

The exquisite spatial resolution of the {\em HST} data reveals for the
first time the presence of two close ($\la$ 1\arcsec) H-band sources
within the cores of 5 objects in our sample: F05024--1941,
F11095--0238, F12072--0444\footnote{Note that the second nucleus in
F12072--0444 was first optically detected in the WFPC2 images of
Surace et al.  (1998) but was eventually classified as a very bright
super stellar cluster based on its location in a magnitude-color
diagram. The spectroscopic results of Dasyra et al. (2005) confirm the
presence of two genuine nuclei in this system.}, F16300$+$1558, and
F21329--2346 (Fig. 3).  The absolute magnitudes ($M_H <$ --21.5) and
disturbed morphologies of these sources seem to rule out the
possibility of bright super star clusters in the hosts or foreground /
background galaxies projected near the source.  We conclude that these
systems are genuine binaries.  In addition, F00456--2904 is a wide
($\sim$ 21\arcsec) binary that was already identified as such from the
ground (VKS02). Our images also often show small galaxies in the
vicinity of the PG~QSOs, but they are considerably fainter
($\Delta$M$_H$ $\ga$ 4 mags) than the QSO hosts.  We have no data to
determine if these small objects are associated or not with the QSOs,
so we list the magnitudes of the objects in the notes to Table 2 but
do not discuss them any further in this paper.  As mentioned in \S
5.3, binary systems are dealt slightly differently from the other
sources, fitting one galaxy component models (PSF + S\'ersic $n$ = free)
for {\em each} nucleus.  Figure 3 shows the results from these fits
and Table 4 lists the linear separation between the nuclei and the
structural parameters for each nucleus.

Table 5 provides a summary of the best-fitting model for each object
in the sample. Binary systems are listed as ``Ambiguous'' in terms of
morphological type, except for F00456--2904 whose primary component is
of a late type and well separated from the companion. The best-fitting
models listed in this table were adopted by inspecting the residuals
and radial plots of Figures 1 -- 3 and the reduced chi-squares,
$\chi_\nu^2$, listed in Tables 2 -- 4. The first of these
$\chi_\nu^2$ values takes into account residuals over the entire
galaxy whereas the second one excludes the central portion which is
affected by errors in the subtraction of the central PSF.  These
reduced chi-squares values should be used with caution when choosing
the best fits. First, we note that they are generally significantly
larger than unity so the fits are not formally very good. This is due
in large part to the presence of merger-induced morphological
anomalies; we return to this important point below (\S 6.2).  We also
notice that the chi-squares tend to be higher for larger, brighter,
and more PSF-dominated objects.  This is not unexpected given the
definition of $\chi_\nu^2$, which is not normalized by the intensity,
and given that the fraction of the detector area that is free of
galaxy emission is more limited for large systems than for small ones.
Thus, $\chi_\nu^2$ cannot be used to compare the goodness of fits
between objects. However, it is a useful quantity to compare the
quality of fits for the same object.

Motivated by what historically has been done, we make a
bimodal choice between preferring $n$ = 1 (late-type) or $n$ = 4
(early-type) and list our choice in Table 5.  To make a bimodal
decision based on a free S\'ersic index $n$, we refer to studies both
published (Barden et al. 2005; McIntosh et al. 2005) and unpublished
(GEMS private communication, Haeussler et al. 2006).  Their findings
are based on computer simulations whereby pure $n$ = 1 and $n$ = 4
galaxies were placed into an image and recovered by fitting.  They can
show that $n$ = 1 galaxies can be recovered quite accurately with a
small scatter in $n$, centered around 1 ($\pm$ 0.5), whereas for $n$ =
4 galaxies the scatter is considerably larger (due to degeneracies
with the sky value and neighboring contamination), but centered around
$n$ = 4.  Therefore, a cut at $n \sim 1.5$ to 2 is one good, somewhat
unbiased, way to decide between galaxies with roughly early or
late-type morphologies.

\subsection{Morphological Type}

A prominent de Vaucouleurs-like component (S\'ersic index $n \approx
4$) is detected in the great majority (22/27 = 81\%) of single-nucleus
systems.  The one galaxy component analysis indicates that a single
spheroidal component often provides a good fit to the surface
brightness distribution of the central portion of these galaxies.  We
see clear trends with {\em IRAS} 25-to-60 $\mu$m colors and optical
spectral types: all warm Seyfert 2s, 1s, and QSOs except
F05189--2524 have a prominent spheroidal component (Fig.  4). This
simple morphological classification is generally consistent on a
object-by-object basis with the ground-based results of Surace et al.
(2001) and VKS02 (see Table 5).

The excellent sensitivity limit of our data allows us to search for
the presence of faint, low-surface-brightness components in all of
these objects.  A faint exponential disk appears to be present in
PG~1119+120 (Bulge/Disk = 2.4), PG~1126--041 (2.3), and PG~1229+204
(3.1) while lobsided or tidally-shredded ``disks'' are visible in
F02021--2103, F05189--2524, F21219--1757, PG0007+106, PG0050+124,
PG0157+001 (Mrk~1014), and PG2130+099. The results of our attempts to
fit this second component as an exponential disk are listed in Table 3
and shown in Figure 2 only for those four cases where the addition of
a second, $n$ = 1 component improved the fit significantly and the
result was physically meaningful (the disk had to be concentric with,
and larger than, the bulge. In F05189--2524, the central ``bulge'' is
very compact and has a steep $n = 1.7$ S\'ersic profile).  We note
that the three QSOs with exponential disks have the lowest infrared
and bolometric luminosities in our sample, bringing further support
for a luminosity dependence of the host morphological type among
quasars (e.g., Dunlop et al. 2003 and references therein).

\subsection{Strength of Tidal Features}

Signs of galactic interactions such as tidal tails and bridges,
lopsided disks, distorted outer isophotes, or double nuclei are
visible in every single system in our sample, including all PG~QSOs.
The residual maps in Figures 1 -- 3 are a particularly good indicator
of these tidal features. In an attempt to quantify the importance of
these features, we first add up the absolute values of the residuals
from the one galaxy component fits over the region unaffected by the
PSF subtraction. Then we normalize this quantity to the total host
luminosities (including tidal features) and call it $R_2$ (see Tables
2 and 3). Although this quantity is sensitive to the presence of
spiral structure, dust lanes, and bright star clusters, we find in our
objects that $R_2$ is dominated by the presence of large-scale
merger-induced anomalies. In Figure 5$a$, we plot $R_2$ versus the
{\em IRAS} 25-to-60 $\mu$m colors for the objects in our sample.  Warm
quasar-like systems tend to have smaller residuals than the other
objects in the sample.  All PG~QSOs and Seyfert 1 galaxies have $R_2
<$ 20\%. Systems with late-type or ambiguous morphologies show larger
residuals than early-type systems (Fig.  5$b$), qualitatively
suggesting that galaxies with a prominent spheroid are in the later
stages of a merger than the late-type and ambiguous systems. This is
consistent with numerical simulations of major (1:1 -- 2:1) mergers
which typically produce elliptical-like galaxies (e.g., Barnes \&
Hernquist 1992; Naab \& Burkert 2001, 2003; Bournaud, Combes, \& Jog
2004; Bournaud, Jog, \& Combes 2005). The spectroscopic results of
Dasyra et al. (2005) also support the major-merger scenario. Note,
however, that the three low-luminosity PG~QSOs in our sample also have
discernible disks; this appears to rule out major mergers for the
origin of these particular systems.

\subsection{Unresolved Nucleus}

An unresolved point source is detected at the centers of most galaxies
in our sample.  This is not surprising given that the galaxies in our
sample were selected on the basis of their high degree of nuclear
concentration (\S 2).  To quantify the importance of the PSF, we
calculate the flux ratio of the PSF to the host, $I_{\rm PSF}$/$I_{\rm
host}$ using the best one galaxy component model for each object.
This ratio is generally less than unity, but varies widely from $\la$
0.01 for F03250+1606 and F04313--1649 to $\sim$ 8 for
F07599+6508. Figure 6 clearly shows that this ratio increases as the
object becomes more AGN-like, either based on its {\em IRAS} 25-to-60
$\mu$m color or its optical spectral type. The largest ratios are
found among Seyfert 1 ULIRGs and PG~QSOs, a sign that the AGN
dominates the central H-band emission in these objects (see
VKS02). This result does not rule out the possibility that a nuclear
starburst is also contributing to the PSF emission, but this starburst
does not produce the bulk of the H-band emission in the nucleus of
Seyfert 1 ULIRGs and PG~QSOs.

\subsection{Host Sizes and Magnitudes} 

Figure 7 shows the distribution of host sizes and absolute magnitudes
for the objects in our sample. The average (median) half-light radii
and H-band absolute magnitudes are 2.68 $\pm$ 1.40 (2.61) kpc and
--24.07 $\pm$ 0.57 (--24.06) mag.\ for the entire sample and 2.59
$\pm$ 1.46 (1.91) kpc and --24.06 $\pm$ 0.56 (--24.05) mag.\ for
singles only. These magnitudes correspond to $\sim$ 1.5 $\pm$ 1
M$^*_H$, where M$^*_H$ = --23.7 mag.\ is the H-band absolute magnitude
of a $L^*$ galaxy in a Schechter function description of the local
field galaxy luminosity function (Cole et al. 2001; VKS02).  A K-S
analysis shows that the hosts of the 7 PG~QSOs in our sample are {\em
  not} significantly different from the hosts of the 1-Jy ULIRGs both
in terms of magnitudes and sizes [P(null) = 0.6 and 0.4,
respectively].

We find good agreement on an object-by-object basis when comparing
these measurements with those of Surace \& Sanders (1999) and Surace
et al. (2001), except for PG~0007+106 where the apparent total
magnitude reported by Surace et al.  (11.81 mag.) is significantly
brighter than ours (13.03 mag., $r \approx$ 4\arcsec) and the 2MASS
magnitude (12.72 mag., $r$ = 7\arcsec).  We suspect an error in the
Surace et al. measurement. The H-band magnitudes of the ULIRG and
PG~QSO hosts in our sample are also quite similar on average with
those measured by McLeod \& McLeod (2001) in other PG~QSOs. We also
have three objects in common with Scoville et al. (2000): F05189-2524,
F07599+6508, and Mrk 1014.  In all three cases, the total magnitudes
listed in this paper are consistent with ours within the uncertaities
of the measurements.

An attempt is made in Figure 8 to compare the R-band measurements from
VKS02 with the H-band measurements listed here. For this exercise, we
correct the R-band magnitudes and surface brightnesses assuming R -- H
= 2.7 mag., a value that is typical for elliptical galaxies at $z
\simeq 0.15$ (Lilly \& Longair 1984; Fukugita et al. 1995). The sizes
measured at R are directly compared with those measured at H. The
hosts extracted from the {\em HST} data are on average smaller by
$\sim$ 2 kpc and fainter by $\sim$ 0.7 mag.\ than the hosts derived
from the ground-based data.  Significant discrepancies in the average
surface brightnesses within half-light radii are also found between
the two data sets although no systematic trends are apparent
(Fig. 8$c$).  Part of the discrepancies in the size and surface
brightness measurements are probably due to radial color gradients
which are well known to be present in these objects (redder colors
near the nucleus; e.g., VKS02). The systematic shift in magnitudes may
indicate an offset in the zero point of the {\em HST} F160W magnitudes
relative to that of the ground-based H-band magnitudes, or perhaps the
assumed elliptical-like R -- H color is too red for the hosts of these
star-forming objects. However, Figure 8$d$ indicates that the
importance of the central PSF is often underestimated in the
ground-based data, especially among Seyfert ULIRGs with bright central
point sources (e.g., F07599+6508). This confirms the suspected
uncertainties associated with the PSF subtraction in the ground-based
measurements of VKS02 for highly nucleated ULIRGs.  The results based
on the {\em HST} data should therefore be used for these particular
objects.  Systems with lower nuclear concentration do not contain
strong nuclear sources so the structural parameters derived by VKS02
for these objects should be reliable.

\section{Discussion}

In \S 1, we posed two important questions: (1) are ULIRGs elliptical
galaxies in formation?  (2) are ULIRGs related to QSOs?  Our {\em HST}
data provide new constraints that can help us revisit these questions.
In this section, we first use the fundamental plane traced by early-type
galaxies (more specifically its projection onto the photometric axes)
to address these issues, and then we attempt to characterize the level
of black-hole driven activity likely to be taking place in the cores
of some of these sources.

\subsection{Fundamental Plane}

Figure 9 shows that ULIRGs with prominent spheroids lie
near the photometric projection of the fundamental plane for
early-type galaxies as traced by the K-band data of Pahre (1999).
However, small ULIRGs are often brighter than inactive
spheroids or bulges of the same size. There are no systematic
differences between the host properties of ULIRGs with Seyfert nuclei
and those with LINER or H~II region-like properties. The excess
emission found in VKS02 in the hosts of Seyfert-1 ULIRGs was due to
under-subtraction of the strong central PSF as discussed in \S 6.4.

Figure 9 also confirms that the hosts of the 7 PG~QSOs in our sample
are {\em not} significantly different from the hosts of the 1-Jy
ULIRGs (see \S 6.4 for more detail).  However, they appear to be
smaller than the hosts of the more luminous QSOs from Dunlop et al
(2003). For this comparison, we directly used the half-light radii
measured from the R-band data of Dunlop et al.  (2003) without
applying any color corrections, while the R-band surface brightness
measurements of Dunlop et al. (2003) were shifted assuming R -- H =
2.9, which is typical for early-type systems at $z$ $\sim$ 0.2 (Lilly
\& Longair 1984; Fukugita, Shimasaku, \& Ichikawa 1995). Negative R --
H radial gradients would reduce the H-band half-light radii, but the
shift between the two sets of QSOs appears too large to be explained
solely by this effect.

Overall, the positions of the ULIRGs and PG~QSOs in our sample are
consistent with those of intermediate-size ($\sim$ 1 -- 2 $L^*$)
ellipticals/lenticulars while the hosts of the luminous QSOs in the
Dunlop et al. (2003) sample are more massive ellipticals. This last
result is not unexpected given the correlation between black hole mass
(and indirectly QSO luminosity) with galaxy mass (see \S 7.2 below).
These results are based purely on photometry and therefore are subject
to errors due to dust extinction and the presence of young stellar
populations in the hosts. The tendency for the ULIRGs with small hosts
to be brighter than spheroids of the same size may be attributable to
excess emission from a young stellar population (Fig. 9).
Nevertheless, it is comforting to see that the conclusions derived
from our photometry are consistent with the kinematic results of Genzel et
al. (2001) and Tacconi et al. (2002) on ULIRGs.

By and large, our results are consistent with the picture where QSO
activity of moderate luminosity is triggered by galaxy mergers that
result in the formation of intermediate-mass spheroids or massive
bulges. The weaker merger-induced morphological anomalies found among
early-type and AGN-like systems (\S 6.2) indicate that QSOs are indeed
preferentially late stage mergers. However, one should remember that
the three lowest luminosity PG~QSOs in our sample also harbor a
significant disk component which is difficult to explain in the
context of major mergers (e.g., Naab \& Burkert 2001, 2003; Bournaud,
Combes, \& Jog 2004; Bournaud, Jog, \& Combes 2005).

\subsection{Level of Black Hole Activity}

The ubiquity of supermassive black holes at the center of galaxies
(e.g., Kormendy \& Gebhardt 2001) suggests that black-hole driven
activity is responsible for at least some of the energy emitted by the
ULIRGs and QSOs in our sample. The host magnitudes derived from our
data can in principle be used to derive the black hole masses in the
cores of these objects, assuming the relation between black hole mass
and the mass of the spheroidal component in normal galaxy (e.g.,
Magorrian et al. 1998; Kormendy \& Gebhardt 2001; Marconi \& Hunt
2003; H\"aring \& Rix 2004) also applies to recent mergers. Using the
H-band early-type host magnitude -- black hole mass relation in
Marconi \& Hunt (2003), log($M_{BH}$) = $-2.80 - (0.464 \times M_H$),
we deduce
black hole masses ranging from $\sim$ 10$^{7.8}$ (F01004--2237 and
F09539+0857) to 10$^{9.0}$ M$_\odot$ (PG0157+001). The average
(median) black hole mass is log($M_{\rm BH}$) = 8.3 $\pm$ 0.3 (8.30).
This derivation once again neglects dust extinction in the hosts
(which would increase $M_{\rm BH}$) and the presence of recent or
on-going star formation in the hosts (which would have the opposite
effect).  These photometric black hole mass estimates are similar on
average with the kinematic estimates of Genzel et al. (2001) and
Tacconi et al. [2002; average (median) black hole mass of log($M_{\rm
BH}$) = 7.9 $\pm$ 0.5 (7.6)].

Next we use these black hole masses and the bolometric luminosities of
these systems to derive their Eddington ratios, {\em i.e.} the
bolometric luminosities expressed in units of their respective
Eddington luminosities.  The bolometric luminosities of ULIRGs are
assumed to be of order 1.15 times their infrared luminosities (e.g.,
Sanders \& Mirabel 1996). The bolometric luminosities of the PG~QSOs
in our sample are calculated using the photometry of Surace et al.
(2001) and Sanders et al. (1989, 1988b). In Figure 10, we see that the
objects in our sample radiate at 5 -- 100\% of the Eddington rate,
assuming {\em all} of the energy emitted by these objects is due to
AGN activity. The values derived for the H~II galaxies should
therefore be considered upper limits since an important fraction of
their energy output probably comes from star formation rather than AGN
activity.  The same is probably also true for ULIRGs with LINER
characteristic as discussed in Veilleux et al.  (1995, 1999a), and
Lutz, Veilleux, \& Genzel (1999). Nevertheless it is interesting to
see that the Eddington ratios measured here are similar to those
measured in quasars (e.g., McLeod \& McLeod 2001 and reference
therein).  In particular, none of the Seyferts and PG~QSOs in our
sample require super-Eddington accretion rates.

\section{Conclusions}

Our detailed two-dimensional analysis of deep {\em HST} NICMOS images
of 26 highly nucleated 1-Jy ULIRGs and 7 PG~QSOs indicates that an
important fraction of them (81\% of the single-nucleus systems)
present early-type morphology well fit by a de Vaucouleurs-like
surface brightness distribution at their centers but with significant
merger-induced morphological anomalies on large scale. This provides
support to the picture where ULIRGs are spheroids in formation. Their
positions relative to the photometric projection of the fundamental
plane for normal (non-active) spheroids are similar to those of
intermediate-size ($\sim$ 1 -- 2 $L^*$) ellipticals or
lenticulars. Indeed, in a few cases, our data reveal a faint disk on
large scale. However, the spheroidal components of small ULIRGs tend
to be brighter than inactive spheroids of the same size. Many of these
results are consistent with the kinematic data of Genzel et al. (2001)
and Tacconi et al. (2002).

The host sizes, magnitudes, and surface brightness distributions of
the 7 PG~QSOs in our sample are statistically identical to those of
the ULIRGs. They all display significant tidal features similar to
those seen in Seyfert ULIRGs, but weaker than those in H~II and LINER
ULIRGs. These results bring further support to the suggestion of a
merger-induced evolutionary connection between ULIRGs and PG~QSOs.
Earlier studies that found a poor match between the host properties of
ULIRGs and those of quasars used a sample of quasars which were
significantly more luminous than the ULIRGs in the 1-Jy sample.
However, the evolutionary sequence may break down at low luminosities:
the disk components found in the lowest luminosity QSOs in our sample
cannot be easily explain in the context of major mergers.

Finally, we add a note of caution when interpreting these
results. The present conclusions are based purely on photometric
measurements which are subject to uncertainties from dust extinction
and young stellar populations. The excess emission from a young
circumnuclear starburst may help explain why ULIRGs with small hosts
are brighter than inactive spheroids. We are in the process of
studying the kinematic properties of the objects in this sample to
test our conclusions.  Mid-infrared spectroscopic data from our
on-going Cycle 1 {\em SST} program will also complement this data set
by providing quantitative measurements of the evolution of the energy
source --- starburst versus AGN --- along the merger sequence. Our
{\em HST} data suggest that AGN-dominated ULIRGs do not require
super-Eddington accretion rates. 

\vskip 0.1in

\acknowledgements We thank the anonymous referee for several
suggestions which improved this paper. S.V., D.C.K., and D.B.S. were
supported in part by NASA through grant GO-0987501A. C.Y.P. is
grateful to STScI for support through the Institute Fellowship
Program. This work has made use of NASA's Astrophysics Data System
Abstract Service and the NASA/IPAC Extragalactic Database (NED), which
is operated by the Jet Propulsion Laboratory, California Institute of
Technology, under contract with the National Aeronautics and Space
Administration.

\clearpage

\begin{deluxetable}{lcccccc}
\tabletypesize{\scriptsize}
\tablenum{1}
\tablewidth{400pt}
\tablecaption{Sample}
\tablehead{
\multicolumn{1}{c}{Name} &
\multicolumn{1}{c}{Other Name} &
\multicolumn{1}{c}{$z$} &
\multicolumn{1}{c}{log$\left(L_{\rm ir} \over L_\odot \right)$} &
\multicolumn{1}{c}{log$\left(L_{\rm bol} \over L_\odot \right)$} &
\multicolumn{1}{c}{$f_{25}/f_{60}$} &
\multicolumn{1}{c}{Spec. Type} \\
\multicolumn{1}{c}{(1)} &
\multicolumn{1}{c}{(2)} &
\multicolumn{1}{c}{(3)} &
\multicolumn{1}{c}{(4)} &
\multicolumn{1}{c}{(5)} &
\multicolumn{1}{c}{(6)} &
\multicolumn{1}{c}{(7)}
}
\startdata
F00188$-$0856    & \nodata &0.128 & 12.33  &  12.39 &  0.143 & LINER     \\
F00397$-$1312    & \nodata &0.261 & 12.90  &  12.96 &  0.180 & HII galaxy\\
F00456$-$2904    & \nodata &0.110 & 12.12  &  12.18 &  0.054 & HII galaxy\\
F01004$-$2237    & \nodata &0.118 & 12.24  &  12.30 &  0.288 & HII galaxy\\
F02021$-$2103    & \nodata &0.116 & 12.01  &  12.07 &  0.207 & \nodata   \\
F03250$+$1606    & \nodata &0.129 & 12.06  &  12.12 &  0.109 & LINER     \\
F04103$-$2838    & \nodata &0.118 & 12.15  &  12.21 &  0.297 & LINER     \\
F04313$-$1649    & \nodata &0.268 & 12.55  &  12.61 &  0.069 & LINER     \\
F05024$-$1941    & \nodata &0.192 & 12.43  &  12.49 &  0.132 & Seyfert 2 \\
F05189$-$2524    & \nodata &0.042 & 12.07  &  12.13 &  0.252 & Seyfert 2 \\
F07599$+$6508    & \nodata &0.149 & 12.46  &  12.52 &  0.314 & Seyfert 1 \\
F09039$+$0503    & \nodata &0.125 & 12.07  &  12.13 &  0.081 & LINER     \\
F09539$+$0857    & \nodata &0.129 & 12.03  &  12.09 &  0.104 & LINER     \\
F11095$-$0238    & \nodata &0.106 & 12.20  &  12.26 &  0.129 & LINER     \\
F11506$+$1331    & \nodata &0.127 & 12.28  &  12.34 &  0.074 & HII galaxy\\
F12072$-$0444    & \nodata &0.129 & 12.35  &  12.41 &  0.220 & Seyfert 2 \\
F12540$+$5708    &  Mrk 231&0.042 & 12.50  &  12.56 &  0.271 & Seyfert 1 \\
F13218$+$0552    & \nodata &0.205 & 12.63  &  12.69 &  0.342 & Seyfert 1 \\
F14070$+$0525    & \nodata &0.265 & 12.76  &  12.82 &  0.131 & Seyfert 2 \\
F14197$+$0813    & \nodata &0.131 & 12.00  &  12.06 &  0.173 & LINER     \\
F15130$-$1958    & \nodata &0.109 & 12.09  &  12.15 &  0.203 & Seyfert 2 \\
F15462$-$0450    & \nodata &0.100 & 12.16  &  12.22 &  0.154 & Seyfert 1 \\
F16300$+$1558    & \nodata &0.242 & 12.63  &  12.69 &  0.047 & LINER     \\
F20414$-$1651    & \nodata &0.086 & 12.14  &  12.20 &  0.080 & HII galaxy\\
F21219$-$1757    & \nodata &0.112 & 12.06  &  12.12 &  0.421 & Seyfert 1 \\
F21329$-$2346    & \nodata &0.125 & 12.09  &  12.15 &  0.073 & LINER     \\
PG0007$+$106$^{(a)}$ & III Zw 2&0.089 & 11.34  &  12.23 &  0.765 &  QSO      \\
PG0050$+$124$^{(a)}$ & I Zw 1  &0.061 & 11.87  &  12.32 &  0.478 &  QSO      \\
PG0157$+$001$^{(a)}$ & Mrk 1014&0.163 & 12.53  &  12.68 &  0.243 &  QSO      \\
PG1119$+$120$^{(a)}$ & Mrk 734 &0.050 & 11.07  &  11.48 &  0.513 &  QSO      \\
PG1126$-$041$^{(a)}$ & Mrk 1298&0.060 & 11.29  &  11.95 &  0.462 &  QSO      \\
PG1229$+$204$^{(a)}$ & Mrk 771 &0.063 & 11.05  &  11.69 &  1.853 &  QSO      \\
PG2130$+$099$^{(a)}$ & Mrk 1513&0.063 & 11.35  &  11.99 &  0.998 &  QSO      \\
\enddata
\tablenotetext{\ }
{{\em Col 1:}\ Object name. PG0157$+$001 is also a 1-Jy ULIRG (F01572$+$0009).}
\tablenotetext{\ }
{{\em Col 2:}\ Other name.}
\tablenotetext{\ }
{{\em Col 3:}\ Redshift from Kim \& Sanders (1998).}
\tablenotetext{\ }
{{\em Col 4:}\ Infrared luminosity calculated using the prescription of Sanders \& Mirabel (1996).}
\tablenotetext{\ }
{{\em Col 5:}\ Bolometric luminosity calculated by assuming that it is
  1.15 times the infrared luminosity of ULIRGs and taken from Surace,
  Sanders, \& Evans (2001) for PG QSOs.}
\tablenotetext{\ }
{{\em Col 6:}\ $f_{25}/f_{60}$ flux ratio. `Warm' objects have
  $f_{25}/f_{60}$ $\ge$ 0.2}
\tablenotetext{\ }
{{\em Col 7:}\ Spectral type from Veilleux, Kim, \& Sanders (1999).}
\tablenotetext{\ } {$^{(a)}$ M$_B$=$-$22.9 and $-$22.0 [PG0007$+$106],
$-$23.5 and $-$22.6 [PG0050$+$124], $-$24.1 and $-$23.2
[PG0157$+$001], $-$22.3 and $-$21.4 [PG1119$+$120], $-$22.9 and
$-$22.0 [PG1126$-$041], $-$22.4 and $-$21.5 [PG1229$+$204], $-$23.3
and $-$22.4 [PG2130$+$099] for H$_0$=50 km s$^{-1}$ Mpc $^{-1}$ and
H$_0$=75 km s$^{-1}$ Mpc $^{-1}$, respectively.}
\end{deluxetable}

\clearpage

\begin{deluxetable}{lrrrrrrrrrrrrr}
\tabletypesize{\scriptsize}
\tablenum{2}
\tablewidth{450pt}
\tablecaption{One Galaxy Component Fits to Single-Nucleus Systems}
\tablehead{
\multicolumn{1}{c}{Name} &
\multicolumn{1}{c}{n} &
\multicolumn{1}{c}{r$_{\frac{1}{2}}$} &
\multicolumn{1}{c}{b/a} &
\multicolumn{1}{c}{PA} &
\multicolumn{1}{c}{m$_n$} &
\multicolumn{1}{c}{M$_n$} &
\multicolumn{1}{c}{m$_{PSF}$} &
\multicolumn{1}{c}{M$_{PSF}$} &
\multicolumn{1}{c}{R$_1$} &
\multicolumn{1}{c}{R$_2$} &
\multicolumn{1}{c}{$\chi^2_{\nu1}$} &
\multicolumn{1}{c}{$\chi^2_{\nu2}$} \\
\multicolumn{1}{c}{(1)} &
\multicolumn{1}{c}{(2)} &
\multicolumn{1}{c}{(3)} &
\multicolumn{1}{c}{(4)} &
\multicolumn{1}{c}{(5)} &
\multicolumn{1}{c}{(6)} &
\multicolumn{1}{c}{(7)} &
\multicolumn{1}{c}{(8)} &
\multicolumn{1}{c}{(9)} &
\multicolumn{1}{c}{(10)} &
\multicolumn{1}{c}{(11)} &
\multicolumn{1}{c}{(12)} &
\multicolumn{1}{c}{(13)}
}
\startdata
F00188$-$0856   &  4.0 &  0.90 &  0.90 &    98.5 &  14.32 & $-$24.23 &    17.84 & $-$20.71 &    9.9 &    7.1 &    4.2 &    3.3 & \\
                &  1.0 &  0.83 &  0.93 &    98.1 &  14.72 & $-$23.83 &    16.53 & $-$22.02 &   39.2 &   30.0 &   18.2 &   13.4 & \\
                &  4.0 &  0.90 &  0.90 &    98.4 &  14.32 & $-$24.23 &    17.83 & $-$20.72 &    9.7 &    6.9 &    4.3 &    3.3 & \\
\\
F00397$-$1312   &  0.8 &  1.89 &  0.65 & $-$25.7 &  15.83 & $-$24.26 &    17.93 & $-$22.16 &   29.0 &   23.5 &    6.7 &    5.5 & \\
                &  1.0 &  1.91 &  0.65 & $-$25.7 &  15.81 & $-$24.28 &    17.95 & $-$22.14 &   27.6 &   23.4 &    6.8 &    5.6 & \\
                &  4.0 &  3.13 &  0.62 & $-$26.1 &  15.53 & $-$24.56 &    18.00 & $-$22.09 &   42.0 &   38.3 &   10.4 &    9.2 & \\
\\
F01004$-$2237   &  5.6 &  0.32 &  0.99 &  $-$4.9 &  15.27 & $-$23.10 &    18.80 & $-$19.57 &    7.2 &    4.6 &    2.3 &    2.2 & \\
                &  1.0 &  0.44 &  0.96 &    26.9 &  15.82 & $-$22.55 &    16.64 & $-$21.73 &   29.9 &   21.3 &    4.6 &    4.0 & \\
                &  4.0 &  0.37 &  0.98 &    24.7 &  15.39 & $-$22.98 &    17.74 & $-$20.63 &    9.7 &    5.8 &    2.4 &    2.2 & \\
\\
F02021$-$2103   &  8.7 & 10.48 &  0.65 &    54.6 &  13.06 & $-$25.27 &    17.86 & $-$20.47 &   22.5 &   20.7 &   16.2 &   16.6 & \\
                &  1.0 &  3.44 &  0.67 &    55.7 &  14.00 & $-$24.33 &    15.90 & $-$22.43 &   42.9 &   31.7 &   49.0 &   31.7 & \\
                &  4.0 &  5.02 &  0.66 &    54.3 &  13.47 & $-$24.86 &    16.54 & $-$21.79 &   24.0 &   18.9 &   19.9 &   16.6 & \\
\\
F03250$+$1606   &  4.1 &  1.66 &  0.73 &  $-$0.6 &  14.10 & $-$24.46 &  \nodata &  \nodata &   13.9 &   13.9 &    5.5 &    5.5 & \\
                &  1.0 &  1.42 &  0.74 &  $-$0.4 &  14.56 & $-$24.00 &    16.60 & $-$21.96 &   39.8 &   35.0 &   19.3 &   15.1 & \\
                &  4.0 &  1.66 &  0.73 &  $-$0.8 &  14.11 & $-$24.45 &    20.09 & $-$18.47 &   13.5 &   13.1 &    5.5 &    5.5 & \\
\\
F04103$-$2838   &  7.4 &  1.72 &  0.81 &     8.3 &  13.97 & $-$24.40 &    17.07 & $-$21.30 &   22.4 &   21.4 &    6.3 &    5.9 & \\
                &  1.0 &  1.54 &  0.84 &    12.0 &  14.68 & $-$23.69 &    15.59 & $-$22.78 &   51.8 &   39.7 &   19.3 &   10.8 & \\
                &  4.0 &  1.51 &  0.81 &     8.8 &  14.22 & $-$24.15 &    16.21 & $-$22.16 &   26.7 &   22.9 &    7.2 &    6.1 & \\
\\
F04313$-$1649   &  4.0 &  3.85 &  0.73 &    84.3 &  15.98 & $-$24.17 &    21.37 & $-$18.78 &   14.7 &   14.6 &    2.9 &    2.9 & \\
                &  1.0 &  2.37 &  0.73 &    83.0 &  16.50 & $-$23.65 &    18.98 & $-$21.17 &   46.3 &   39.2 &    4.0 &    3.8 & \\
                &  4.0 &  3.77 &  0.74 &    85.0 &  15.97 & $-$24.18 &    21.30 & $-$18.85 &   16.0 &   15.7 &    2.9 &    2.9 & \\
\\
F05189$-$2524   &  7.6 &  0.28 &  0.94 &    75.2 &  11.92 & $-$24.21 &    12.76 & $-$23.37 &   23.0 &   11.0 &   66.2 &   27.3 & \\
                &  1.0 &  0.41 &  0.93 &    82.2 &  12.57 & $-$23.56 &    12.26 & $-$23.87 &   47.2 &   28.7 &  154.1 &  110.3 & \\
                &  4.0 &  0.35 &  0.94 &    75.9 &  12.16 & $-$23.97 &    12.49 & $-$23.64 &   27.9 &   15.2 &   73.5 &   40.1 & \\
\\
F07599$+$6508   & 20.0 &  4.72 &  0.96 &    72.8 &  13.53 & $-$25.35 &    12.14 & $-$26.74 &   61.9 &    9.2 &   27.9 &    4.8 & \\
                &  1.0 &  3.90 &  0.95 &    51.0 &  14.69 & $-$24.19 &    12.08 & $-$26.80 &  102.9 &   20.2 &   34.1 &    5.9 & \\
                &  4.0 &  4.04 &  0.96 &    50.7 &  14.20 & $-$24.68 &    12.09 & $-$26.79 &   83.9 &   11.1 &   30.3 &    4.5 & \\
\\
F09039$+$0503   &  1.4 &  1.61 &  0.73 & $-$21.0 &  14.77 & $-$23.72 &    18.38 & $-$20.11 &   34.0 &   32.5 &   14.0 &   13.4 & \\
                &  1.0 &  1.51 &  0.73 & $-$21.8 &  14.85 & $-$23.64 &    18.14 & $-$20.35 &   37.9 &   35.5 &   15.1 &   14.6 & \\
                &  4.0 &  3.01 &  0.73 & $-$19.4 &  14.40 & $-$24.09 &    19.11 & $-$19.38 &   33.9 &   32.3 &   16.9 &   16.5 & \\
\\
F09539$+$0857   &  3.3 &  1.19 &  0.82 &   140.9 &  15.35 & $-$23.21 &    17.25 & $-$21.31 &   25.6 &   21.3 &    6.8 &    6.4 & \\
                &  1.0 &  1.17 &  0.84 &   139.8 &  15.81 & $-$22.75 &    17.25 & $-$21.31 &   44.6 &   33.2 &   13.4 &    8.2 & \\
                &  4.0 &  2.61 &  0.85 &   141.5 &  15.35 & $-$23.21 &    17.25 & $-$21.31 &   45.4 &   29.2 &   13.7 &    7.0 & \\
\\
F11506$+$1331   &  3.6 &  1.35 &  0.79 & $-$78.2 &  14.66 & $-$23.87 &    18.76 & $-$19.77 &   25.8 &   22.7 &   12.1 &   11.7 & \\
                &  1.0 &  1.05 &  0.84 & $-$62.8 &  15.05 & $-$23.48 &    17.06 & $-$21.47 &   40.9 &   33.7 &   16.9 &   15.4 & \\
                &  4.0 &  1.40 &  0.78 & $-$78.7 &  14.62 & $-$23.91 &    19.30 & $-$19.23 &   25.3 &   23.1 &   12.1 &   12.0 & \\
\\
F12540$+$5708   & 11.5 &  1.38 &  0.93 &   179.4 &  11.14 & $-$24.99 &    10.57 & $-$25.56 &   41.5 &    8.1 &  148.8 &   86.3 & \\
                &  1.0 &  1.69 &  0.89 &   170.6 &  12.05 & $-$24.08 &    10.43 & $-$25.70 &   62.1 &   14.2 &  297.0 &  106.5 & \\
                &  4.0 &  1.31 &  0.92 &   178.6 &  11.57 & $-$24.56 &    10.49 & $-$25.64 &   35.8 &    8.8 &  164.2 &   74.0 & \\
\\
F13218$+$0552   &  5.5 &  4.56 &  0.86 &$-$173.1 &  14.83 & $-$24.74 &    14.62 & $-$24.95 &   27.1 &   13.0 &    5.5 &    3.6 & \\
                &  1.0 &  3.44 &  0.87 &   175.5 &  15.43 & $-$24.14 &    14.53 & $-$25.04 &   49.9 &   31.5 &    7.4 &    3.9 & \\
                &  4.0 &  4.12 &  0.86 &$-$173.7 &  14.95 & $-$24.62 &    14.60 & $-$24.97 &   27.1 &   14.6 &    5.6 &    3.5 & \\
\\
F14070$+$0525   &  4.2 &  3.58 &  0.83 & $-$20.6 &  15.40 & $-$24.73 &    19.31 & $-$20.82 &   19.6 &   18.9 &    5.1 &    4.5 & \\
                &  1.0 &  2.67 &  0.82 & $-$22.3 &  15.86 & $-$24.27 &    18.08 & $-$22.05 &   37.5 &   34.5 &    7.0 &    4.6 & \\
                &  4.0 &  3.38 &  0.83 & $-$20.4 &  15.43 & $-$24.70 &    19.31 & $-$20.82 &   20.0 &   19.3 &    4.6 &    4.5 & \\
\\
F14197$+$0813   &  2.0 &  2.00 &  0.88 &   176.2 &  14.57 & $-$24.03 &    17.43 & $-$21.17 &   20.2 &   17.5 &   12.1 &   10.2 & \\
                &  1.0 &  1.69 &  0.91 &   178.3 &  14.78 & $-$23.82 &    17.12 & $-$21.48 &   31.6 &   28.3 &   15.5 &   13.0 & \\
                &  4.0 &  2.72 &  0.89 &   176.3 &  14.32 & $-$24.28 &    18.23 & $-$20.37 &   20.2 &   15.2 &   14.3 &   12.0 & \\
\\
F15130$-$1958   & 12.7 &  2.00 &  0.78 & $-$78.2 &  14.14 & $-$24.06 &    15.70 & $-$22.50 &   17.5 &   14.2 &    8.0 &    6.2 & \\
                &  1.0 &  1.71 &  0.74 & $-$80.0 &  15.09 & $-$23.11 &    15.04 & $-$23.16 &   38.3 &   28.0 &   13.9 &    6.6 & \\
                &  4.0 &  1.51 &  0.77 & $-$79.3 &  14.64 & $-$23.56 &    15.27 & $-$22.93 &   21.5 &   18.9 &    8.9 &    5.9 & \\
\\
F15462$-$0450   & 12.2 & 24.42 &  0.86 &$-$119.0 &  13.45 & $-$24.56 &    14.24 & $-$23.77 &   27.1 &   17.0 &    9.8 &    6.4 & \\
                &  1.0 &  3.38 &  0.72 &$-$118.9 &  14.81 & $-$23.20 &    14.12 & $-$23.89 &   40.3 &   23.0 &   13.0 &    7.3 & \\
                &  4.0 &  5.20 &  0.83 &$-$118.5 &  14.25 & $-$23.76 &    14.18 & $-$23.83 &   28.8 &   15.5 &    9.6 &    6.1 & \\
\\
F20414$-$1651   &  1.7 &  1.42 &  0.43 &$-$122.7 &  14.41 & $-$23.27 &    17.32 & $-$20.36 &   24.8 &   19.7 &   12.6 &    8.5 & \\
                &  1.0 &  1.28 &  0.43 &$-$121.9 &  14.52 & $-$23.16 &    17.53 & $-$20.15 &   30.9 &   26.0 &   13.8 &   11.9 & \\
                &  4.0 &  2.10 &  0.43 &$-$123.4 &  14.20 & $-$23.48 &    17.32 & $-$20.36 &   25.5 &   18.3 &   20.2 &   13.0 & \\
\\
F21219$-$1757   & 11.1 &  8.76 &  0.84 &   130.1 &  12.93 & $-$25.33 &    13.27 & $-$24.99 &   21.0 &    6.8 &   14.8 &    5.7 & \\
                &  1.0 &  3.10 &  0.91 &   103.8 &  13.96 & $-$24.30 &    13.12 & $-$25.14 &   43.1 &   14.5 &   31.7 &   11.2 & \\
                &  4.0 &  3.91 &  0.86 &   129.4 &  13.46 & $-$24.80 &    13.19 & $-$25.07 &   23.9 &    6.4 &   16.1 &    5.6 & \\
\\
PG0007$+$106$^{(a)}$    &  5.6 &  4.22 &  0.90 &   140.3 &  13.62 & $-$24.14 &    13.64 & $-$24.12 &   28.5 &   15.2 &    9.7 &    4.0 & \\
                &  1.0 &  2.49 &  0.91 &   131.4 &  14.32 & $-$23.44 &    13.56 & $-$24.20 &   52.1 &   32.9 &   18.8 &    9.0 & \\
                &  4.0 &  3.42 &  0.90 &   139.9 &  13.79 & $-$23.97 &    13.62 & $-$24.14 &   30.0 &   16.2 &    9.8 &    4.0 & \\
\\
PG0050$+$124    &  4.0 &  1.84 &  0.83 &    36.7 &  12.33 & $-$24.61 &    12.24 & $-$24.70 &   36.0 &    9.5 &  106.8 &   14.0 & \\
                &  1.0 &  1.85 &  0.85 &    34.1 &  12.79 & $-$24.15 &    12.14 & $-$24.80 &   52.6 &   15.9 &  159.0 &   33.2 & \\
                &  4.0 &  1.84 &  0.83 &    36.6 &  12.34 & $-$24.60 &    12.24 & $-$24.70 &   36.0 &    9.5 &  106.8 &   14.0 & \\
\\
PG0157$+$001$^{(a)}$    & 17.0 &  5.47 &  0.93 & $-$40.0 &  13.06 & $-$26.01 &    14.08 & $-$24.99 &   22.2 &   13.4 &   14.5 &    8.6 & \\
                &  1.0 &  2.89 &  0.90 &   111.1 &  14.26 & $-$24.81 &    13.65 & $-$25.42 &   56.5 &   35.0 &   38.6 &   19.2 & \\
                &  4.0 &  2.80 &  0.94 & $-$50.2 &  13.74 & $-$25.33 &    13.77 & $-$25.30 &   32.1 &   19.5 &   17.9 &   10.0 & \\
\\
PG1119$+$120$^{(a)}$    &  4.5 &  1.36 &  0.88 &   166.6 &  12.91 & $-$23.60 &    13.34 & $-$23.17 &   18.9 &    8.3 &   19.7 &    9.6 & \\
                &  1.0 &  1.08 &  0.99 &   160.5 &  13.46 & $-$23.05 &    13.21 & $-$23.30 &   43.5 &   24.1 &   47.1 &   24.0 & \\
                &  4.0 &  1.30 &  0.89 &   165.9 &  12.96 & $-$23.55 &    13.32 & $-$23.19 &   18.9 &    8.1 &   19.6 &    9.7 & \\
\\
PG1126$-$041$^{(a)}$    &  3.0 &  3.44 &  0.41 &   151.5 &  12.97 & $-$23.93 &    12.35 & $-$24.55 &   39.6 &    6.6 &   59.0 &    9.9 & \\
                &  1.0 &  2.83 &  0.41 &   152.6 &  13.35 & $-$23.55 &    12.32 & $-$24.58 &   52.4 &   14.0 &   71.6 &   18.2 & \\
                &  4.0 &  3.93 &  0.41 &   151.4 &  12.85 & $-$24.05 &    12.36 & $-$24.54 &   38.8 &    6.3 &   59.6 &   10.6 & \\
\\
PG1229$+$204    &  4.3 &  5.05 &  0.72 &    27.1 &  12.62 & $-$24.39 &    13.72 & $-$23.29 &   17.9 &   13.0 &   14.7 &   12.1 & \\
                &  1.0 &  3.34 &  0.66 &    28.6 &  13.20 & $-$23.81 &    13.59 & $-$23.42 &   31.4 &   18.1 &   47.0 &   25.9 & \\
                &  4.0 &  4.78 &  0.72 &    27.2 &  12.66 & $-$24.35 &    13.71 & $-$23.30 &   18.0 &   12.9 &   14.7 &   12.0 & \\
\\
PG2130$+$099$^{(a)}$    &  9.8 & 14.52 &  0.62 &$-$131.9 &  12.31 & $-$24.70 &    12.42 & $-$24.59 &   19.3 &    6.3 &   20.0 &   11.7 & \\
                &  1.0 &  2.92 &  0.61 &$-$135.0 &  13.54 & $-$23.47 &    12.36 & $-$24.65 &   32.8 &   11.1 &   38.3 &   19.7 & \\
                &  4.0 &  4.14 &  0.62 &$-$132.6 &  13.01 & $-$24.00 &    12.40 & $-$24.61 &   21.4 &    5.7 &   20.8 &   11.6 & \\
\\
\enddata
\tablenotetext{\ }
{{\it Col 1:}\ Object name.}
\tablenotetext{\ }
{{\it Col 2:}\ S\'ersic index.}
\tablenotetext{\ }
{{\it Col 3:}\ Half-light radius in kpc of S\'ersic component.}
\tablenotetext{\ }
{{\it Col 4:}\ Axis ratio of S\'ersic component.}
\tablenotetext{\ }
{{\it Col 5:}\ Position angle of major axis of S\'ersic component.}
\tablenotetext{\ }
{{\it Col 6:}\ Apparent H magnitude of S\'ersic component.}
\tablenotetext{\ }
{{\it Col 7:}\ Absolute H magnitude of S\'ersic component.}
\tablenotetext{\ }
{{\it Col 8:}\ Apparent H magnitude of PSF component.}
\tablenotetext{\ }
{{\it Col 9:}\ Absolute H magnitude of PSF component.}
\tablenotetext{\ }
{{\it Col 10:}\ Absolute residuals normalized to total host galaxy flux (\%).}
\tablenotetext{\ }
{{\it Col 11:}\ PSF-masked absolute residuals normalized to total host galaxy
flux (\%). The central PSF region brighter than 10 H mag arcsec$^{-2}$ was
masked for these calculations.}
\tablenotetext{\ }
{{\it Col 12:}\ Reduced $\chi^2$ value.}
\tablenotetext{\ }
{{\it Col 13:}\ PSF-masked reduced $\chi^2$ value. The central PSF region  
brighter than 10 H mag arcsec$^{-2}$ was masked for these calculations.}
\tablenotetext{\ }
{}
\tablenotetext{\ }
{$^{(a)}$ Faint galaxies are detected in the vicinity of these QSOs.
  The H-band absolute magnitudes of these galaxies are $-20.09$
  (PG0007+106), $-20.02$ (PG0157+001), $-19.66$ (PG1119+120), $-19.66$
  (PG1126$-$041), and $-19.66$ (PG2130+099). }
\end{deluxetable}

\clearpage

\begin{center}
\begin{deluxetable}{lrrrrrrrrrrrrr}
\tabletypesize{\scriptsize}
\tablenum{3}
\tablewidth{465pt}
\tablecaption{Two Galaxy Component Fits to Single-Nucleus Systems}
\tablehead{
\multicolumn{1}{c}{Name} &
\multicolumn{1}{c}{n} &
\multicolumn{1}{c}{r$_{\frac{1}{2}}$} &
\multicolumn{1}{c}{b/a} &
\multicolumn{1}{c}{PA} &
\multicolumn{1}{c}{m$_n$} &
\multicolumn{1}{c}{M$_n$} &
\multicolumn{1}{c}{m$_{PSF}$} &
\multicolumn{1}{c}{M$_{PSF}$} &
\multicolumn{1}{c}{R$_1$} &
\multicolumn{1}{c}{R$_2$} &
\multicolumn{1}{c}{$\chi^2_{\nu1}$} &
\multicolumn{1}{c}{$\chi^2_{\nu2}$} \\
\multicolumn{1}{c}{(1)} &
\multicolumn{1}{c}{(2)} &
\multicolumn{1}{c}{(3)} &
\multicolumn{1}{c}{(4)} &
\multicolumn{1}{c}{(5)} &
\multicolumn{1}{c}{(6)} &
\multicolumn{1}{c}{(7)} &
\multicolumn{1}{c}{(8)} &
\multicolumn{1}{c}{(9)} &
\multicolumn{1}{c}{(10)} &
\multicolumn{1}{c}{(11)} &
\multicolumn{1}{c}{(12)} &
\multicolumn{1}{c}{(13)} 
}
\startdata
PG1119$+$120 &  1.0 & 3.52 & 0.41 & $-$19.3 &  14.28 & -22.23 &    13.37 & $-$23.14 &   15.2 &    5.7 &   14.4 &    4.1 & \\
             &  4.0 & 0.71 & 0.87 &$-$111.8 &  13.32 & -23.19 &          &          &        &        &        &        & \\
PG1126$-$041 &  1.0 & 5.44 & 0.38 & $-$10.2 &  14.26 & -22.64 &    12.36 & $-$24.54 &   37.4 &    5.5 &   57.5 &    8.2 & \\
             &  4.0 & 2.54 & 0.38 &   152.4 &  13.35 & -23.55 &          &          &        &        &        &        & \\
PG1229$+$204 &  1.0 & 6.03 & 0.27 &   124.2 &  14.35 & -22.66 &    13.73 & $-$23.28 &   13.8 &    9.8 &   10.8 &    8.9 & \\
             &  4.0 & 2.70 & 0.91 & $-$10.8 &  13.11 & -23.90 &          &          &        &        &        &        & \\
\\
\hline
\\
F05189$-$2524$^{(a)}$&  1.7 & 0.29 & 0.94 &    81.1 &  12.46 & -23.67 &    12.39 & $-$23.74 &    9.1 &    3.1 &   34.2 &    8.3 & \\
             &  0.8 & 3.88 & 0.91 &    52.2 &  13.29 & -22.84 &          &          &        &        &        &        & \\
\enddata
\tablenotetext{\ }
{{\it Col 1:}\ Object name.}
\tablenotetext{\ }
{{\it Col 2:}\ S\'ersic index.}
\tablenotetext{\ }
{{\it Col 3:}\ Half-light radius in kpc of S\'ersic component.}
\tablenotetext{\ }
{{\it Col 4:}\ Axis ratio of S\'ersic component.}
\tablenotetext{\ }
{{\it Col 5:}\ Position angle of major axis of S\'ersic component.}
\tablenotetext{\ }
{{\it Col 6:}\ Apparent H magnitude of S\'ersic component.}
\tablenotetext{\ }
{{\it Col 7:}\ Absolute H magnitude of S\'ersic component.}
\tablenotetext{\ }
{{\it Col 8:}\ Apparent H magnitude of PSF component.}
\tablenotetext{\ }
{{\it Col 9:}\ Absolute H magnitude of PSF component.}
\tablenotetext{\ }
{{\it Col 10:}\ Absolute residuals normalized to total host galaxy flux (\%).}
\tablenotetext{\ }
{{\it Col 11:}\ PSF-masked absolute residuals normalized to total host galaxy
flux (\%). The central PSF region brighter than 10 H mag arcsec$^{-2}$ was
masked for these calculations.}
\tablenotetext{\ }
{{\it Col 12:}\ Reduced $\chi^2$ value.}
\tablenotetext{\ }
{{\it Col 13:}\ PSF-masked reduced $\chi^2$ value. The central PSF region  
brighter than 10 H mag arcsec$^{-2}$ was masked for these calculations.}
\tablenotetext{\ }
{}
\tablenotetext{\ }
{$^{(a)}$ Note that a very compact ``bulge'' with $n$ = 1.7 is
  detected in this object. This object is classified as late type in
  the text and all of the figures.}
\end{deluxetable}
\end{center}

\clearpage

\begin{deluxetable}{lccrrrrrrrrrrrrr}
\tabletypesize{\scriptsize}
\tablenum{4}
\tablewidth{515pt}
\tablecaption{One Galaxy Component Fits to Binary Systems}
\tablehead{
\multicolumn{1}{c}{Name} &
\multicolumn{1}{c}{} &
\multicolumn{1}{c}{NS} &
\multicolumn{1}{c}{n} &
\multicolumn{1}{c}{r$_{\frac{1}{2}}$} &
\multicolumn{1}{c}{b/a} &
\multicolumn{1}{c}{PA} &
\multicolumn{1}{c}{m$_n$} &
\multicolumn{1}{c}{M$_n$} &
\multicolumn{1}{c}{m$_{PSF}$} &
\multicolumn{1}{c}{M$_{PSF}$} &
\multicolumn{1}{c}{R$_1$} &
\multicolumn{1}{c}{R$_2$} &
\multicolumn{1}{c}{$\chi^2_{\nu1}$} &
\multicolumn{1}{c}{$\chi^2_{\nu2}$} \\
\multicolumn{1}{c}{(1)} &
\multicolumn{1}{c}{} &
\multicolumn{1}{c}{(2)} &
\multicolumn{1}{c}{(3)} &
\multicolumn{1}{c}{(4)} &
\multicolumn{1}{c}{(5)} &
\multicolumn{1}{c}{(6)} &
\multicolumn{1}{c}{(7)} &
\multicolumn{1}{c}{(8)} &
\multicolumn{1}{c}{(9)} &
\multicolumn{1}{c}{(10)} &
\multicolumn{1}{c}{(11)} &
\multicolumn{1}{c}{(12)} &
\multicolumn{1}{c}{(13)} &
\multicolumn{1}{c}{(14)}
}
\startdata
F00456$-$2904 & NE &20.7 &  5.1 & 5.52 & 0.84 & $-$91.5 & 16.23 &$-$21.99 &  21.35 &$-$16.87 &   22.5 &   21.1 &    0.8 &    0.7 & \\
              & SW &     &  1.8 & 1.85 & 0.94 &   101.0 & 14.02 &$-$24.20 &  16.74 &$-$21.48 &   17.3 &   16.8 &   11.6 &   10.3 & \\
F05024$-$1941 & NE & 3.0 & 17.8 & 5.49 & 0.82 & $-$89.7 & 14.78 &$-$24.65 &  17.75 &$-$21.68 &   11.7 &    9.8 &    4.0 &    3.6 & \\
              & SW &     &  1.8 & 3.20 & 0.45 & $-$79.1 & 15.61 &$-$23.82 &  20.43 &$-$19.00 &        &        &        &        & \\
F11095$-$0238 & NE & 0.9 &  2.3 & 1.67 & 0.72 &$-$150.9 & 15.59 &$-$22.55 &  18.52 &$-$19.62 &   22.5 &   20.1 &    4.3 &    4.2 & \\
              & SW &     &  3.7 & 1.45 & 0.53 &$-$162.9 & 16.25 &$-$21.89 &  18.89 &$-$19.25 &        &        &        &        & \\
F12072$-$0444 & N  & 2.0 & 14.7 &10.88 & 0.80 &$-$154.4 & 14.76 &$-$23.80 &  18.04 &$-$20.52 &   22.5 &   19.1 &    8.6 &    6.4 & \\
              & S  &     &  0.4 & 3.58 & 0.79 &$-$172.9 & 14.75 &$-$23.81 &  17.14 &$-$21.42 &        &        &        &        & \\
F16300$+$1558 & N  & 4.4 &  1.2 & 3.50 & 0.38 &$-$42.2  & 16.70 &$-$23.23 &  22.11 &$-$17.82 &   24.2 &   23.8 &    4.2 &    4.2 & \\
              & S  &     &  2.2 & 4.27 & 0.73 &   150.4 & 15.19 &$-$24.74 &  20.06 &$-$19.87 &        &        &        &        & \\
F21329$-$2346 & NE & 2.4 &  3.5 & 2.67 & 0.60 &   172.4 & 15.03 &$-$23.46 &  19.39 &$-$19.10 &   14.2 &   13.8 &    4.8 &    4.7 & \\
              & SW &     &  1.2 & 0.59 & 0.77 &   119.9 & 16.96 &$-$21.53 &  20.88 &$-$17.61 &        &        &        &        & \\
\enddata
\tablenotetext{\ }
{{\it Col 1:}\ Object name.}
\tablenotetext{\ }
{{\it Col 2:}\ Nuclear Separation in kpc.}
\tablenotetext{\ }
{{\it Col 3:}\ S\'ersic index.}
\tablenotetext{\ }
{{\it Col 4:}\ Half-light radius in kpc of S\'ersic component.}
\tablenotetext{\ }
{{\it Col 5:}\ Axis ratio of S\'ersic component.}
\tablenotetext{\ }
{{\it Col 6:}\ Position angle of major axis of S\'ersic component.}
\tablenotetext{\ }
{{\it Col 7:}\ Apparent H magnitude of S\'ersic component.}
\tablenotetext{\ }
{{\it Col 8:}\ Absolute H magnitude of S\'ersic component.}
\tablenotetext{\ }
{{\it Col 9:}\ Apparent H magnitude of PSF component.}
\tablenotetext{\ }
{{\it Col 10:}\ Absolute H magnitude of PSF component.}
\tablenotetext{\ }
{{\it Col 11:}\ Absolute residuals normalized to total host galaxy flux (\%).}
\tablenotetext{\ }
{{\it Col 12:}\ PSF-masked absolute residuals normalized to total host galaxy
flux (\%). The central PSF region brighter than 10 H mag arcsec$^{-2}$ was
masked for these calculations.}
\tablenotetext{\ }
{{\it Col 13:}\ Reduced $\chi^2$ value.}
\tablenotetext{\ }
{{\it Col 14:}\ PSF-masked reduced $\chi^2$ value. The central PSF region 
brighter than 10 H mag arcsec$^{-2}$ was masked for these calculations.}
\end{deluxetable}

\clearpage

\begin{deluxetable}{llrrcccccccccc}
\tabletypesize{\tiny}
\tablenum{5}
\tablecaption{Summary$^{(a)}$}
\tablehead{
\multicolumn{1}{c}{Name} &
\multicolumn{1}{l}{}&
\multicolumn{1}{c}{M$_{PSF}$} &
\multicolumn{1}{c}{M$_{host}$} &
\multicolumn{1}{c}{M$_{model}$} &
\multicolumn{1}{c}{M$_{total}$} &
\multicolumn{1}{c}{${\rm I}_{PSF} \over {\rm I}_{host}$} &
\multicolumn{1}{c}{${\rm I}_{model} \over {\rm I}_{host}$} &
\multicolumn{1}{c}{r$_{\frac{1}{2}}$} &
\multicolumn{1}{c}{$<\mu_{\frac{1}{2}}>$} &
\multicolumn{1}{c}{MC} &
\multicolumn{1}{c}{MC-V02} &
\multicolumn{1}{c}{IC} \\
\multicolumn{1}{c}{(1)} &
\multicolumn{1}{l}{}&
\multicolumn{1}{c}{(2)} &
\multicolumn{1}{c}{(3)} &
\multicolumn{1}{c}{(4)} &
\multicolumn{1}{c}{(5)} &
\multicolumn{1}{c}{(6)} &
\multicolumn{1}{c}{(7)} &
\multicolumn{1}{c}{(8)} &
\multicolumn{1}{c}{(9)} &
\multicolumn{1}{c}{(10)} &
\multicolumn{1}{c}{(11)} &
\multicolumn{1}{c}{(12)}
}
\startdata
F00188$-$0856&   &$-$20.72 &$-$24.19 &$-$24.23 &$-$24.23 &   0.04 &   1.04 &   0.90 &  14.60 &       E &       E &       V & \\
F00397$-$1312&   &$-$22.14 &$-$24.35 &$-$24.28 &$-$24.48 &   0.13 &   0.94 &   1.91 &  16.43 &       A &     E/D &       V & \\
F00456$-$2904&   &\nodata  &\nodata  &\nodata  &\nodata  &\nodata & \nodata&\nodata &\nodata &    L, A & \nodata &    IIIa & \\
             &NE &$-$16.87 &$-$21.80 &$-$21.99 &$-$21.81 &   0.00 &   1.19 &   5.26 &  20.63 &       A & \nodata & \nodata & \\
             &SW &$-$21.48 &$-$24.23 &$-$24.20 &$-$24.31 &   0.08 &   0.97 &   1.76 &  16.32 &       L &       E & \nodata & \\
F01004$-$2237&   &$-$20.63 &$-$23.02 &$-$22.98 &$-$23.13 &   0.11 &   0.96 &   0.37 &  14.27 &       E &     E/D &       V & \\
F02021$-$2103&   &$-$21.79 &$-$24.72 &$-$24.86 &$-$24.79 &   0.07 &   1.14 &   5.02 &  17.09 &    E+L? &       E &     IVa & \\
F03250$+$1606&   &$-$18.47 &$-$24.42 &$-$24.45 &$-$24.42 &   0.00 &   1.03 &   1.66 &  15.41 &       E &       E &     IVb & \\
F04103$-$2838&   &$-$22.16 &$-$24.25 &$-$24.15 &$-$24.40 &   0.15 &   0.91 &   1.51 &  15.56 &       E &       A &     IVb & \\
F04313$-$1649&   &$-$18.85 &$-$23.99 &$-$24.18 &$-$24.00 &   0.01 &   1.19 &   3.77 &  17.85 &       E &     E/D &     IVa & \\
F05024$-$1941&   &$-$23.30 &$-$24.49 &$-$24.38 &$-$24.80 &   0.33 &   0.90 &   3.76 &  17.63 &    A, A &     E/D &    IIIb & \\
             &NE &$-$21.68 &\nodata  &$-$24.65 &\nodata  &\nodata & \nodata&   5.49 &  18.18 &       A & \nodata & \nodata & \\
             &SW &$-$19.00 &\nodata  &$-$23.82 &\nodata  &\nodata & \nodata&   3.20 &  17.84 &       A & \nodata & \nodata & \\
F05189$-$2524&   &$-$23.74&$-$23.96 &$-$24.09 &$-$24.62 &   0.82 &   0.89 &   0.53 &  13.23 &       L &       E &     IVb & \\
F07599$+$6508&   &$-$26.79 &$-$24.55 &$-$24.68 &$-$26.92 &   7.87 &   1.13 &   4.04 &  17.30 &       E &       E &     IVb & \\
F09039$+$0503&   &$-$20.35 &$-$24.05 &$-$23.64 &$-$24.09 &   0.03 &   0.69 &   1.51 &  16.19 &       L &       A &     IVa & \\
F09539$+$0857&   &$-$21.31 &$-$22.92 &$-$23.21 &$-$23.14 &   0.23 &   1.31 &   2.61 &  17.76 &       E &     E/D &       V & \\
F11095$-$0238&   &$-$20.26 &$-$23.01 &$-$22.81 &$-$23.09 &   0.08 &   0.83 &   1.65 &  17.18 &    A, A &       A &    IIIb & \\
             &NE &$-$19.62 &\nodata  &$-$22.55 &\nodata  &\nodata & \nodata&   1.67 &  17.46 &       A & \nodata & \nodata & \\
             &SW &$-$19.25 &\nodata  &$-$21.89 &\nodata  &\nodata & \nodata&   1.45 &  17.80 &       A & \nodata & \nodata & \\
F11506$+$1331&   &$-$19.23 &$-$23.97 &$-$23.91 &$-$23.98 &   0.01 &   0.95 &   1.40 &  15.62 &       E &       A &     IVb & \\
F12072$-$0444&   &$-$22.25 &$-$24.22 &$-$24.23 &$-$24.38 &   0.16 &   1.01 &   3.65 &  17.53 &    A, A &       E &    IIIb & \\
             &N  &$-$20.52 &\nodata  &$-$23.80 &\nodata  &\nodata & \nodata&   3.58 &  17.91 &       A & \nodata & \nodata & \\
             &S  &$-$21.42 &\nodata  &$-$23.81 &\nodata  &\nodata & \nodata&  10.87 &  20.34 &       A & \nodata & \nodata & \\
F12540$+$5708&   &$-$25.64 &$-$24.52 &$-$24.56 &$-$25.97 &   2.81 &   1.04 &   1.31 &  14.65 &       E &       E &     IVb & \\
F13218$+$0552&   &$-$24.97 &$-$24.47 &$-$24.62 &$-$25.50 &   1.58 &   1.15 &   4.12 &  17.50 &       E &       E &       V & \\
F14070$+$0525&   &$-$20.82 &$-$24.62 &$-$24.70 &$-$24.65 &   0.03 &   1.08 &   3.38 &  17.22 &       E &     E/D &       V & \\
F14197$+$0813&   &$-$21.48 &$-$24.01 &$-$23.82 &$-$24.11 &   0.10 &   0.84 &   1.69 &  16.29 &       A &       A &       V & \\
F15130$-$1958&   &$-$22.93 &$-$23.62 &$-$23.56 &$-$24.08 &   0.53 &   0.95 &   1.51 &  16.05 &       E &       E &     IVb & \\
F15462$-$0450&   &$-$23.83 &$-$23.40 &$-$23.76 &$-$24.39 &   1.49 &   1.39 &   5.20 &  18.44 &       E &     E/D &     IVb & \\
F16300$+$1558&   &$-$21.08 &$-$24.96 &$-$24.80 &$-$24.99 &   0.03 &   0.86 &   4.10 &  17.52 &       A &     E/D &    IIIb & \\
             &N  &$-$17.82 &\nodata  &$-$23.23 &\nodata  &\nodata & \nodata&   3.50 &  18.75 &       A & \nodata & \nodata & \\
             &S  &$-$19.87 &\nodata  &$-$24.74 &\nodata  &\nodata & \nodata&   4.27 &  17.67 &       A & \nodata & \nodata & \\
F20414$-$1651&   &$-$20.15 &$-$23.34 &$-$23.16 &$-$23.40 &   0.05 &   0.85 &   1.28 &  16.21 &       A &     E/D &     IVb & \\
F21219$-$1757&   &$-$25.07 &$-$24.59 &$-$24.80 &$-$25.61 &   1.56 &   1.21 &   3.91 &  16.89 &    E+L? &       E &       V & \\
F21329$-$2346&   &$-$19.33 &$-$23.62 &$-$23.72 &$-$23.64 &   0.02 &   1.10 &   3.65 &  17.52 &    A, A &       E &    IIIb & \\
             &NE &$-$19.10 &\nodata  &$-$23.46 &\nodata  &\nodata & \nodata&   2.67 &  17.61 &       A & \nodata & \nodata & \\
             &SW &$-$17.61 &\nodata  &$-$21.53 &\nodata  &\nodata & \nodata&   0.59 &  16.28 &       A & \nodata & \nodata & \\
PG0007$+$106 &   &$-$24.14 &$-$23.79 &$-$23.97 &$-$24.73 &   1.38 &   1.18 &   3.42 &  17.40 &       E & \nodata &       V & \\
PG0050$+$124 &   &$-$24.70 &$-$24.53 &$-$24.60 &$-$25.37 &   1.17 &   1.07 &   1.84 &  15.27 &       E & \nodata &       V & \\
PG0157$+$001 &   &$-$25.30 &$-$25.39 &$-$25.33 &$-$26.10 &   0.92 &   0.95 &   2.80 &  15.94 &       E &       E &     IVb & \\
PG1119$+$120 &   &$-$23.19 &$-$23.44 &$-$23.55 &$-$24.07 &   0.79 &   1.11 &   1.30 &  15.62 &     E+L & \nodata &       V & \\
PG1126$-$041 &   &$-$24.54 &$-$23.79 &$-$24.05 &$-$24.98 &   2.00 &   1.27 &   3.93 &  16.76 &     E+L & \nodata &       V & \\
PG1229$+$204 &   &$-$23.30 &$-$24.06 &$-$24.35 &$-$24.50 &   0.50 &   1.31 &   4.78 &  17.39 &     E+L & \nodata &       V & \\
PG2130$+$099 &   &$-$24.61 &$-$23.76 &$-$24.00 &$-$25.02 &   2.19 &   1.25 &   4.14 &  17.30 &    E+L? & \nodata &       V & \\
\enddata
\tablenotetext{\ }
{{\it Col 1:}\ Object name.}
\tablenotetext{\ }
{{\it Col 2:}\ Absolute magnitude of PSF component.}
\tablenotetext{\ }
{{\it Col 3:}\ Absolute magnitude of host galaxy (including tidal features).}
\tablenotetext{\ }
{{\it Col 4:}\ Absolute magnitude of best-fitting galaxy host model.}
\tablenotetext{\ }
{{\it Col 5:}\ Total absolute magnitude (host + PSF).}
\tablenotetext{\ }
{{\it Col 6:}\ PSF-to-host intensity ratio.}
\tablenotetext{\ }
{{\it Col 7:}\ Model-to-host intensity ratio.}
\tablenotetext{\ }
{{\it Col 8:}\ Half-light radius in kpc of S\'ersic component.}
\tablenotetext{\ }
{{\it Col 9:}\ Mean surface brightness within half-light radius in 
H mag.\ arcsec$^{-2}$.}
\tablenotetext{\ }
{{\it Col 10:}\ Morphological class: E = early type, L = late type, A
= ambiguous. ``E + L'' indicates a galaxy with a dominant early-type
component surrounded by an exponential disk. Question marks (``?'')
indicate uncertain classifications. The components in the binaries are
classified as ambiguous by default, unless the components are widely
separated as in the case of F00456--2904. }
\tablenotetext{\ }
{{\it Col 11:}\ Morphological class from VKS02.}
\tablenotetext{\ }
{{\it Col 12:}\ Revised interaction class based on new HST results
(see VSK02 for details on the definitions).}  
\tablenotetext{\ }
{}
\tablenotetext{\ }
{$^{(a)}$ Entries in this table are the parameters from the
  best-fitting one galaxy component models for single objects (Table
  2) and for binary systems (Table 4). As explained in \S 6, a bimodal
  choice was made between preferring $n = 1$ (late type) or $n = 4$
  (early type). The only exception is F05189--2524, where the
  parameters from the two galaxy component fits (Table 3) were used
  for the calculations since the one galaxy component fit is a very
  poor representation of the data. The magnitudes and half-light
  radius of this galaxy were calculated from the sum of both
  components listed in Table 3. Unless otherwise noted, the figures in
  the paper used the entries in this table. }
\end{deluxetable}


\begin{figure*}[ht]
\caption{ Results from the GALFIT one galaxy component analysis of
  single-nucleus systems. For each object, panel ($a$) shows the
  original data while the lower panels show the residuals after
  subtracting three different models: ($b$) PSF + S\'ersic component
  with unconstrained index, ($c$) PSF + S\'ersic component with $n$ =
  1 (exponential disk), and ($d$) PSF + S\'ersic component with $n$ =
  4 (de Vaucouleurs spheroid). The intensity scale is logarithmic and
  the horizontal segment in panel ($a$) represents 10 kpc. The
  tickmarks in panels (a), (b), (c), and (d) are separated by
  5\arcsec. Panel ($e$) compares the observed radial surface
  brightness profile with those of the models: solid black line is the
  data, short-dashed red line is PSF + $n$ = free, long-dashed green
  line is PSF + $n$ = 1, and dotted blue line is PSF + $n$ = 4. The
  various PSFs used for these models are omitted for clarity. The
  bottom section of panel ($e$) shows the residual intensities after
  subtracting the models from the data. }
\end{figure*}
\clearpage
 
\begin{figure*}[ht]
\epsscale{0.7}
\caption{ Results from the GALFIT two galaxy component analysis of
  single-nucleus systems. A low-surface-brightness exponential disk is
  detected unambiguously in only four systems. Systems with lopsided
  or tidally-shredded disks are not shown here because the
  two galaxy component analysis does not provide significantly better fits
  than the one galaxy component analysis for these objects.  Panel ($a$)
  shows the original data and panel ($b$) shows the residuals after
  subtracting a model with a PSF, a bulge-like S\'ersic
  component with $n$ = 4, and a disk-like S\'ersic component with $n$
  = 1. In the case of F05189--2524, the ``bulge-like'' component has
  $n$ = 1.7 and the disk-like component has $n$ = 0.8. Panels ($c$)
  and ($d$) show the surface brightness distributions of the two
  S\'ersic components used in the model. The centroids of the
  components are left unconstrained. The intensity scale is
  logarithmic and the vertical segment between panels ($b$) and ($c$)
  represents 10 kpc. The tickmarks in each panel are separated by
  5\arcsec.}
\end{figure*}

\begin{figure*}[ht]
\epsscale{1.05}
\caption{ Results from the GALFIT analysis for binary systems.  Panels
  on the left show the original data and panels on the right show the
  residuals after subtracting a model that assumes the sum of a PSF
  component and a S\'ersic component with unconstrained centroid and
  S\'ersic index for each nucleus in the system.  Table 4 lists the
  parameters for each of these models. The intensity scale is
  logarithmic and the vertical segment between the panels represents
  10 kpc. The tickmarks in each panel are separated by 5\arcsec. }
\end{figure*}

\begin{figure*}[ht]
\epsscale{1.05}
\plotone{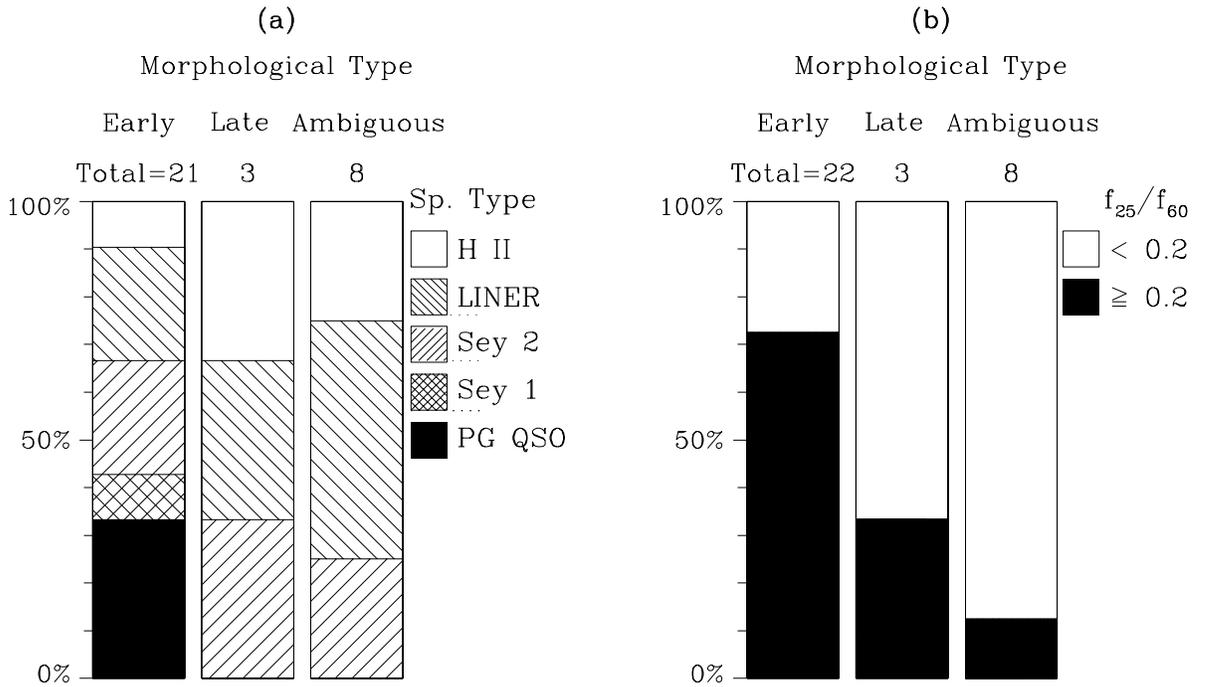}
\caption{ Trends between the dominant morphological types based on the
  one galaxy component decomposition and ($a$) optical spectral types,
  ($b$) {\em IRAS} 25-to-60 $\mu$m colors. There is one fewer object
  in panel ($a$) than in panel ($b$) because the optical spectral type
  of F02021$-$2103 is unknown. The hosts of warm, quasar-like objects all
  have a prominent early-type spheroidal component.  F05189--2524 is
  the only Seyfert 2 ULIRG in the sample with a dominant late-type
  morphology.}
\end{figure*}

\begin{figure*}[ht]
\epsscale{1.0}
\plotone{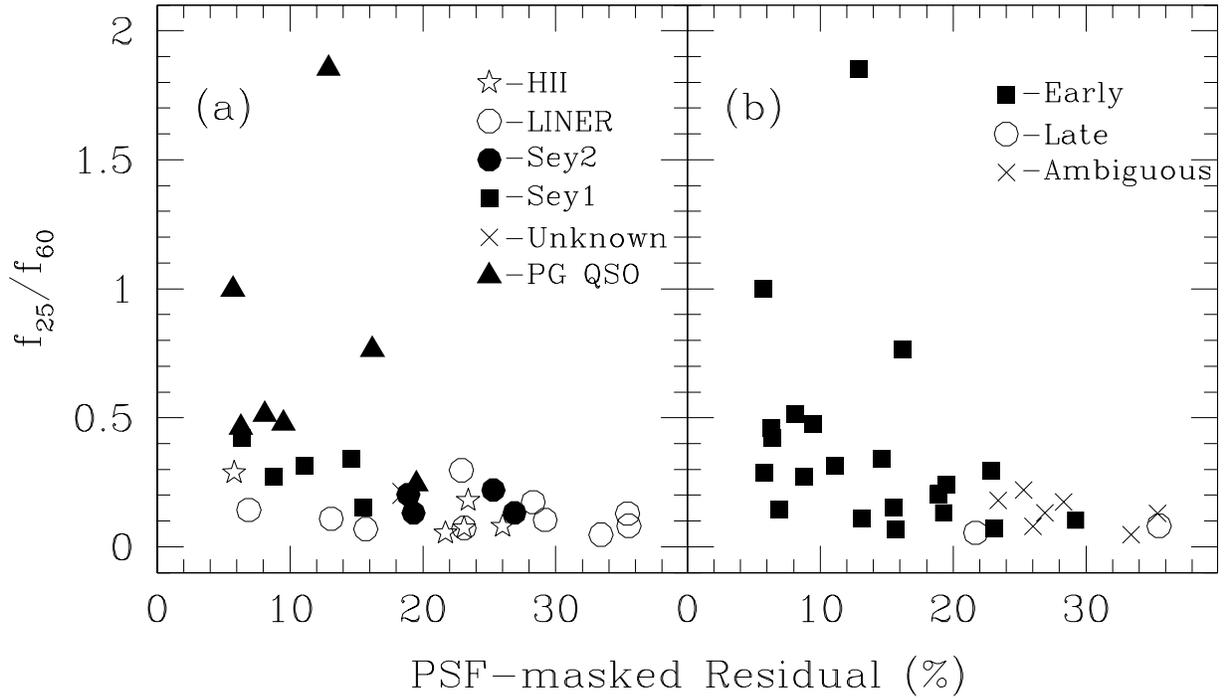}
\caption{ {\em IRAS} 25-to-60 $\mu$m colors plotted against the
  PSF-masked residuals as defined in the text. The residuals are
  calculated from the one galaxy component fits. F05189--2524 is not
  shown in these panels because a one galaxy component fit is a very
  poor representation of the data (see Fig. 1). The residuals are
  smaller among warm, quasar-like objects ($a$) with dominant
  early-type morphology ($b$).  However, note in ($b$) that some of
  the quasars classified as having a dominant early-type morphology
  also harbour faint exponential disks on largel scale.}
\end{figure*}

\begin{figure*}[ht]
\epsscale{1.1}
\plotone{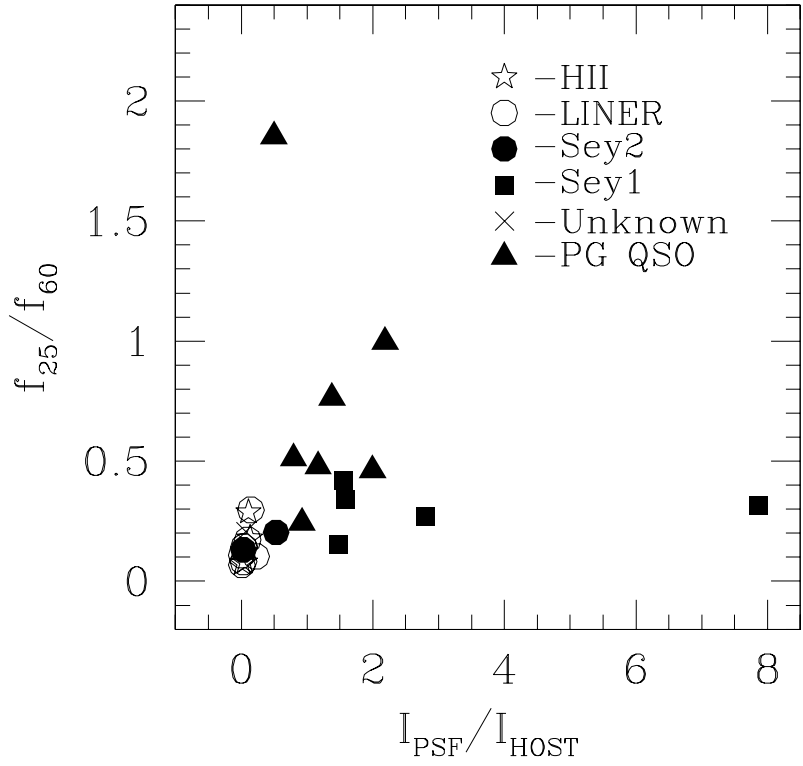}
\caption{ {\em IRAS} 25-to-60 $\mu$m colors plotted against $I_{\rm
    PSF}/I_{\rm host}$, the intensity of the PSF component normalized
  to that of the host galaxy. The object with $f_{25}/f_{100} = 1.853$
  is PG1229+204, while the object with $I_{\rm PSF}/I_{\rm host}
  \simeq 8 $ is F07599+6508. Warm, quasar-like objects have stronger
  PSF components than H~II and LINER ULIRGs. }
\end{figure*}

\begin{figure*}[ht]
\epsscale{1.0}
\plotone{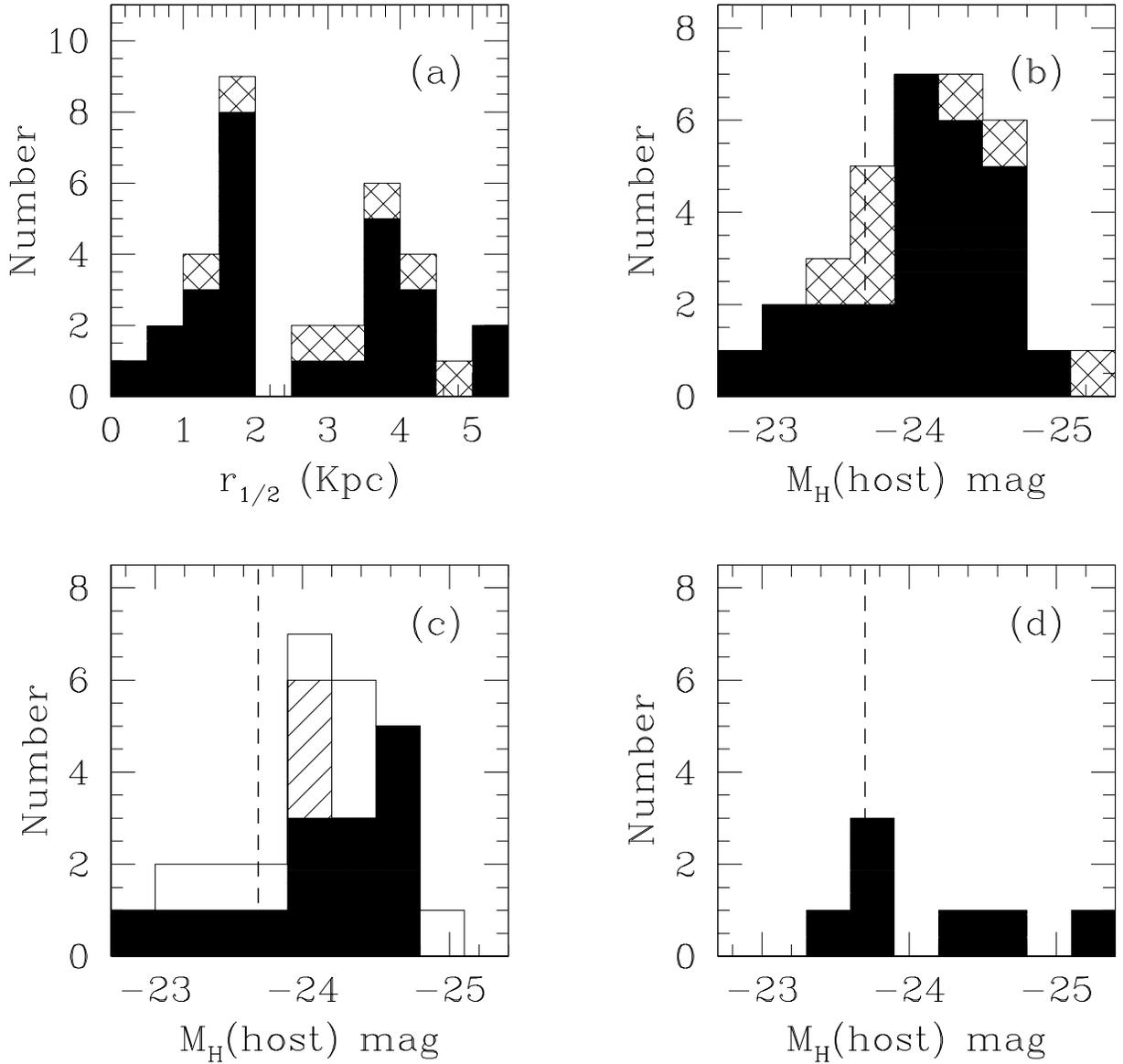}
\caption{ Distribution of the half-light radii and host absolute
  magnitudes for the objects in the {\em HST} sample.  In panels ($a$)
  and ($b$), the ULIRGs are in black and the PG~QSOs are
  cross-hatched. A K-S test on these data indicates no significant
  difference between the host sizes and magnitudes of the 1-Jy ULIRGs
  and PG~QSOs in this sample.  Panel ($c$) shows the distribution of
  host absolute magnitudes for ULIRGs according to their morphology
  (black represents dominant early type, hatched corresponds to late
  type, and white indicates ambiguous systems). Panel ($d$) is the
  same as panel ($c$) but for the PG~QSOs. No obvious trends with
  dominant morphological type are seen in the data. The vertical
  dashed line in panels ($b$), ($c$), and ($d$) represents $M^*_H =
  -23.7$ mags, the H-band absolute magnitude of a $L^*$ galaxy in a
  Schechter function description of the local field galaxy luminosity
  function.  }
\end{figure*}

\begin{figure*}[ht]
\epsscale{1.0}
\plotone{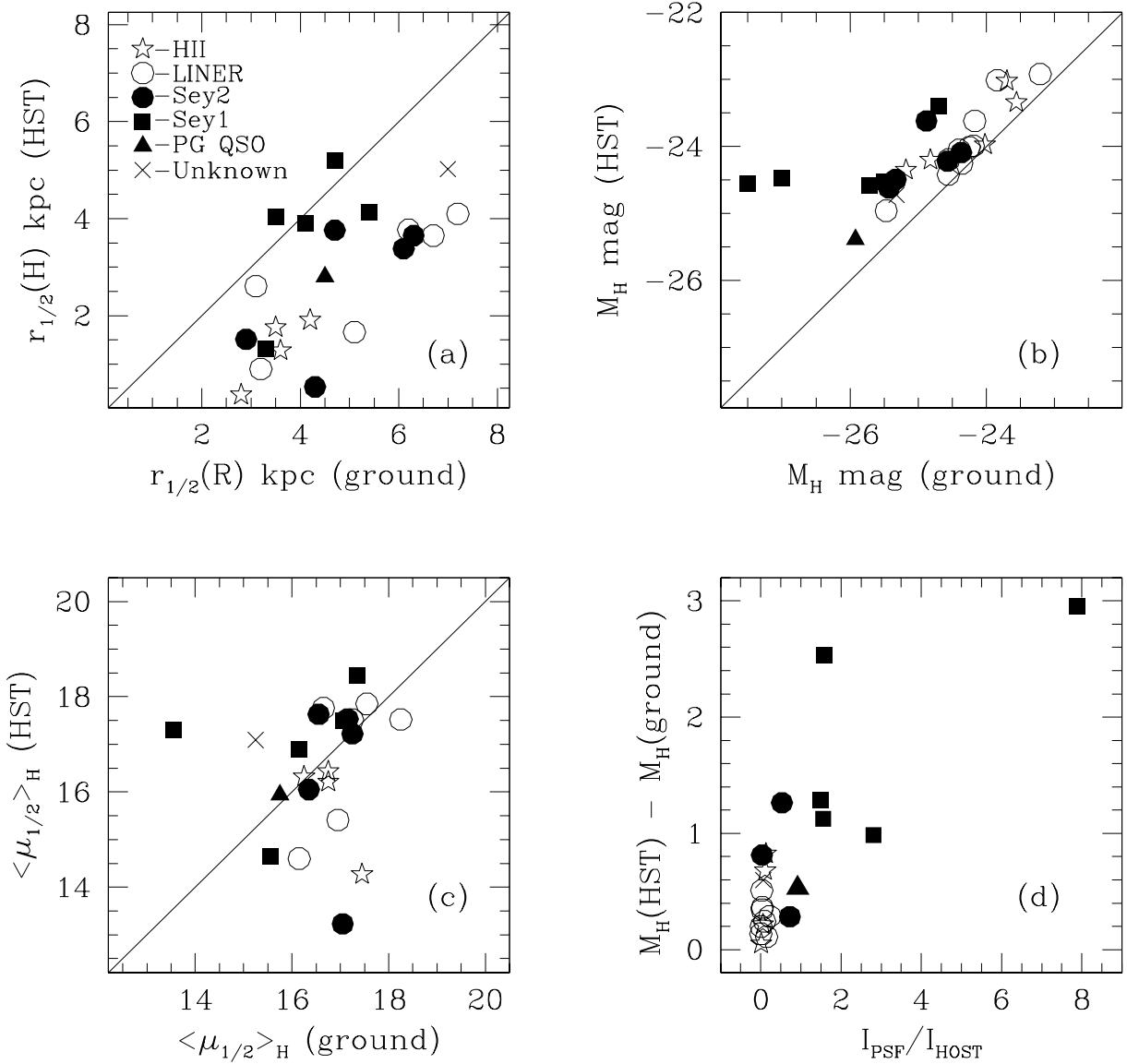}
\caption{ Attempt to compare the results from the present H-band study
  with those from the R-band study of VKS02. ($a$) half-light radii,
  ($b$) host absolute magnitudes, ($c$) average surface brightnesses
  within half-light radius. Part of the discrepancies in host sizes
  may be due to radial color gradients (see text). A color typical of
  elliptical galaxies at $z \simeq 0.15$ is assumed (R -- H = 2.7
  mag.) when comparing the magnitudes and surface brightnesses from
  the two studies. Some of the discrepancies in surface brightnesses
  and magnitudes may be due in part to this choice of color. However,
  panel ($d$) shows that the discrepancies in the magnitude
  measurements scale with the strength of the PSF component, therefore
  suggesting that the PSF component in the R-band data has been
  under-subtracted.  }
\end{figure*}
 
\begin{figure*}[ht]
\epsscale{0.95}
\plotone{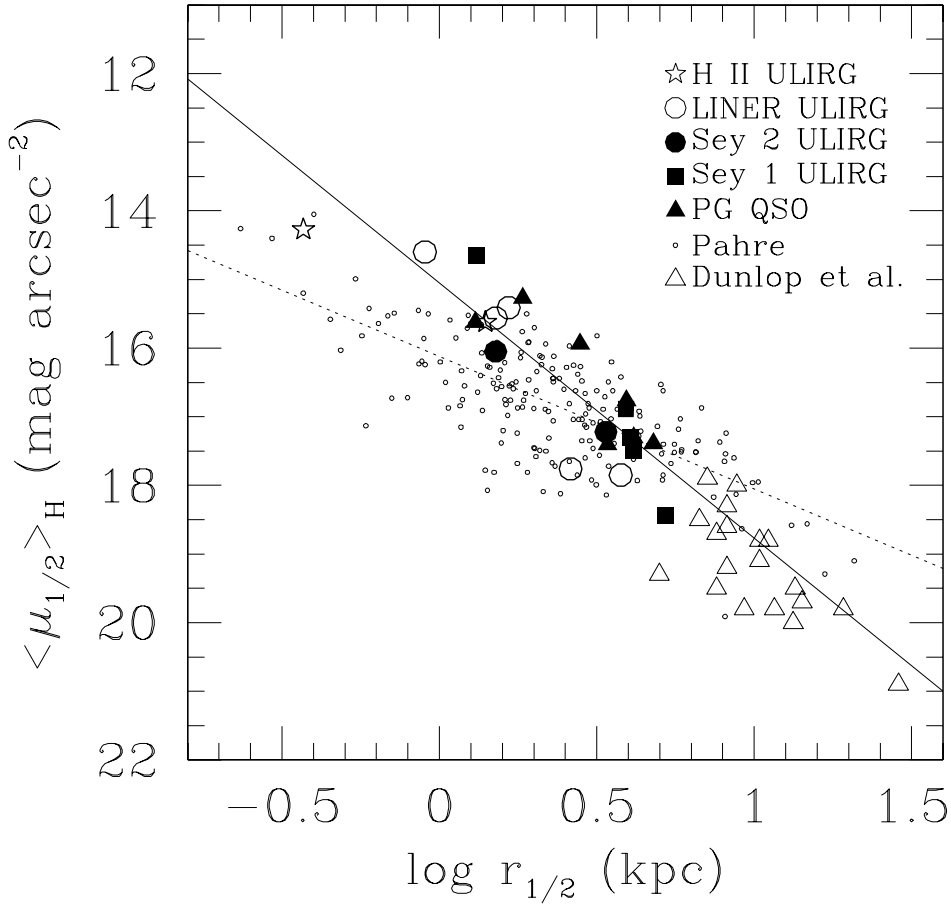}
\caption{ Surface brightnesses versus half-light radii for the
  early-type host galaxies in the {\em HST} sample.  The hosts of the
  7 PG~QSOs in our sample are statistically indistinguishable from the
  hosts of the 1-Jy ULIRGs. Both classes of objects fall near the
  photometric fundamental plane relation of ellipticals as traced by
  the data of Pahre (1999; dashed line), although the smaller objects
  in our sample tend to lie above this relation (the solid line is a
  linear fit through our data points).  This may be due to excess
  H-band emission from a young stellar population.  ULIRGs and PG~QSOs
  populate the region of the photometric fundamental plane of
  intermediate-size ($\sim$ 1 -- 2 $L^*$) elliptical/lenticular
  galaxies.  In contrast, the hosts of the luminous quasars of Dunlop
  et al.  (2003) are massive ellipticals which are significantly
  larger than the hosts of ULIRGs and PG~QSOs.  For this comparison,
  the R-band half-light radii tabulated in Dunlop et al. were taken at
  face value, and the surface brightnesses in that paper were shifted
  assuming R -- H = 2.9, which is typical for early-type systems at
  $z$ $\sim$ 0.2 (see text for more detail).}
\end{figure*}

\begin{figure*}[ht]
\epsscale{1.1}
\plotone{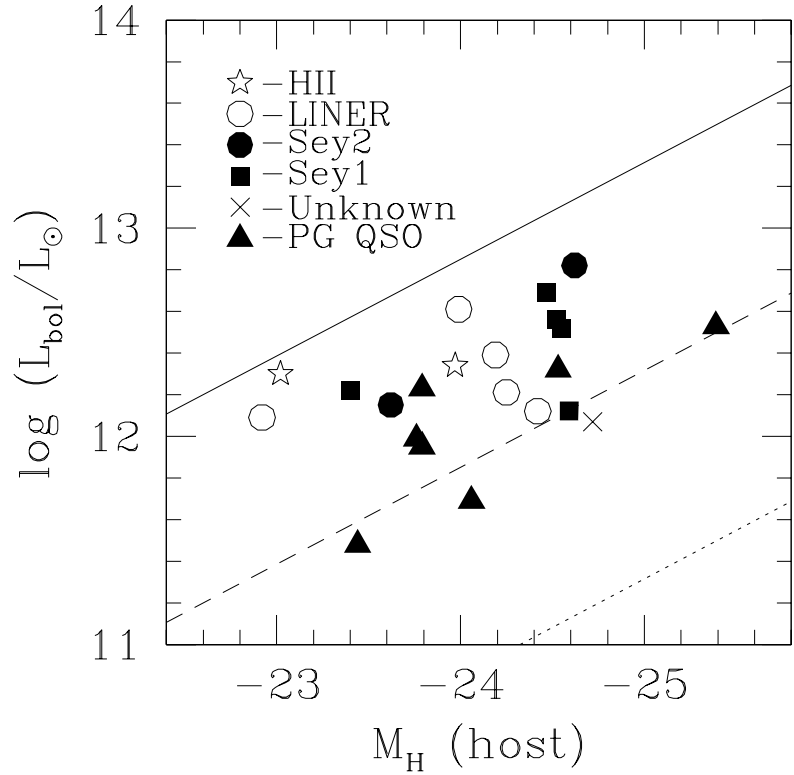}
\caption{ Total bolometric luminosities of ULIRGs and PG~QSOs with
  dominant early-type hosts versus absolute magnitudes of the
  spheroidal components. Diagonal dotted, dashed, and solid lines
  represent 1\%, 10\%, 100\% of the Eddington luminosity using the
  relation of Marconi \& Hunt (2003) to translate spheroid magnitudes
  into black hole masses.  The Eddington ratios derived from this
  figure should be considered upper limits since starbursts may
  contribute an important fraction of the energy in some of these
  objects, especially those classified as H~II galaxies and LINERs.
  None of the objects in the sample radiate at super-Eddington rates.}
\end{figure*}

\end{document}